\documentclass[aps,prd,twocolumn,floatfix,nofootinbib,superscriptaddress,tightenlines]{revtex4}

\usepackage[dvipsnames]{xcolor}
\usepackage{amsmath}
\usepackage{dcolumn}
\usepackage{lipsum}
\usepackage{amssymb}
\usepackage{url}
\usepackage{epsfig}
\usepackage{graphicx}
\usepackage{amsmath}
\usepackage{bm}
\usepackage{setspace}
\usepackage{lscape}
\usepackage{amsthm}
\usepackage{bbold}
\usepackage{dcolumn}
\usepackage{epsfig}
\usepackage{graphics}
\usepackage{graphicx}
\usepackage{longtable}
\usepackage{svg}

\usepackage{bm}
\usepackage{xspace}
\usepackage{cancel}
\usepackage{float}
\usepackage{multirow}
\definecolor{darkgreen}{rgb}{0,0.5,0}
\definecolor{purple}{rgb}{0.5,0,0.5}
\definecolor{nblue}{rgb}{0.0,0.0,0.50}
\definecolor{scarlet}{rgb}{1.0,0.2,0}
\definecolor{darkmagenta}{rgb}{0.55, 0.0, 0.55}
\definecolor{darkolivegreen}{rgb}{0.33, 0.42, 0.18}
\definecolor{darkcandyapplered}{rgb}{0.64, 0.0, 0.0}
\usepackage[colorlinks=true, pdfstartview=FitV, linkcolor=purple, citecolor= purple, urlcolor=blue]{hyperref}
\usepackage{microtype}

\def\qon#1{q_{#1,0}^{(+)}}

\def\bra#1{\langle{#1}|}
\def\ket#1{|{#1}\rangle}

\def\beq{\begin{equation}} \def\eeq{\end{equation}}
\def\beqn{\begin{eqnarray}} \def\eeqn{\end{eqnarray}}
 
\def\nn{\nonumber}
\def\Eq#1{Eq.~(\ref{#1})}

\newcommand{\la}{\langle}
\newcommand{\ra}{\rangle}

\def\ket#1{|{#1}\ra}
\def\bra#1{\la{#1}|}
\def\ketbra#1#2{|{#1}\ra \la{#2}|}

\newcommand{\lan}{\langle}
\newcommand{\ran}{\rangle}

\def\qon#1{q_{#1,0}^{(+)}}

\newcommand{\expval}[1]{\lan {#1} \ran}

\pdfoutput=1

\DeclareMathOperator{\Ker}{Ker}

\begin{document}
\title{Variational quantum eigensolver for causal loop Feynman diagrams and directed acyclic graphs}

\author{Giuseppe Clemente}
\email{giuseppe.clemente@desy.de}
\affiliation{Deutsches Elektronen-Synchrotron DESY, Platanenallee 6, 15738 Zeuthen, Germany.}

\author{Arianna Crippa}
\email{arianna.crippa@desy.de}
\affiliation{Deutsches Elektronen-Synchrotron DESY, Platanenallee 6, 15738 Zeuthen, Germany.} 

\author{Karl Jansen}
\email{karl.jansen@desy.de}
\affiliation{Deutsches Elektronen-Synchrotron DESY, Platanenallee 6, 15738 Zeuthen, Germany.}

\author{Selomit Ram\'irez-Uribe}
\email{selomit@ific.uv.es}
\affiliation{Instituto de F\'{\i}sica Corpuscular, Universitat de Val\`{e}ncia -- Consejo Superior de Investigaciones Cient\'{\i}ficas, Parc Cient\'{\i}fic, E-46980 Paterna, Valencia, Spain.}
\affiliation{Facultad de Ciencias F\'{\i}sico-Matem\'aticas,
Universidad Aut\'onoma de Sinaloa, Ciudad Universitaria, CP 80000 Culiac\'an, Mexico.}
\affiliation{Facultad de Ciencias de la Tierra y el Espacio,
Universidad Aut\'onoma de Sinaloa, Ciudad Universitaria, CP 80000 Culiac\'an, Mexico.}

\author{Andr\'es E. Renter\'ia-Olivo}
\email{andres.renteria@ific.uv.es}
\affiliation{Instituto de F\'{\i}sica Corpuscular, Universitat de Val\`{e}ncia -- Consejo Superior de Investigaciones Cient\'{\i}ficas, Parc Cient\'{\i}fic, E-46980 Paterna, Valencia, Spain.}

\author{Germ\'an Rodrigo}
\email{german.rodrigo@csic.es}
\affiliation{Instituto de F\'{\i}sica Corpuscular, Universitat de Val\`{e}ncia -- Consejo Superior de Investigaciones Cient\'{\i}ficas, Parc Cient\'{\i}fic, E-46980 Paterna, Valencia, Spain.}

\author{German F. R. Sborlini}
\email{german.sborlini@usal.es}
\affiliation{Departamento de F\'isica Fundamental e IUFFyM, Universidad de Salamanca, 
37008 Salamanca, Spain.}
\affiliation{Escuela de Ciencias, Ingenier\'ia y Diseño, Universidad Europea de Valencia, \\ Paseo de la Alameda 7, 46010 Valencia, Spain.}

\author{Luiz Vale Silva}
\email{luizva@ific.uv.es}
\affiliation{Instituto de F\'{\i}sica Corpuscular, Universitat de Val\`{e}ncia -- Consejo Superior de Investigaciones Cient\'{\i}ficas, Parc Cient\'{\i}fic, E-46980 Paterna, Valencia, Spain.}

\begin{abstract}
We present a variational quantum eigensolver (VQE) algorithm for the efficient bootstrapping of the causal representation of multiloop Feynman diagrams in the Loop-Tree Duality (LTD) 
or, equivalently, the selection of acyclic configurations 
in directed graphs. A loop Hamiltonian based on the adjacency matrix 
describing a multiloop topology, and whose different energy levels 
correspond to the number of cycles, is minimized by VQE to 
identify the causal or acyclic configurations. The algorithm has been adapted to select multiple degenerated minima and thus achieves higher detection rates. A performance comparison with a Grover's based algorithm is discussed in detail. The VQE approach requires, in general, fewer qubits and shorter circuits for its implementation, albeit with lesser success rates. 

\end{abstract}

\maketitle               

%\tableofcontents

%%%%%%%%%%%%%%%%%%%%%%%%%%%%%%%%%%
%%%%%%%%%%%%%%%%%%%%%%%%%%%%%%%%%%%%%%%%%%%%%%%%%%%%%%%%%%%%%%%%%%%%%%%%%%%%%%
\section{Introduction}
\label{sec:introduction}
In recent years, there has been a tremendous progress in achieving highly-precise theoretical predictions for particle colliders, which has been possible because of the development of new techniques in Quantum Field Theories (QFT) as well as improved (classical) hardware. For instance, from the theory point of view, the calculation of multiloop scattering amplitudes and Feynman integrals was optimized by applying different sophisticated techniques. Some of these theoretical advancements include Mellin-Barnes transformations~\cite{Blumlein:2000hw,Bierenbaum:2006mq,Gluza:2007rt,Freitas:2010nx,Dubovyk:2016ocz}, algebraic reduction of integrands~\cite{Mastrolia:2011pr,Badger:2012dp,Zhang:2012ce,Mastrolia:2012an,Mastrolia:2012wf,Ita:2015tya,Mastrolia:2016dhn,Ossola:2006us}, integration-by-parts identities~\cite{Chetyrkin:1981qh,Laporta:2001dd}, contour deformation assisted by neural networks~\cite{Winterhalder:2021ngy} and sector decomposition~\cite{Binoth:2000ps,Smirnov:2008py,Carter:2010hi,Borowka:2017idc}, among several other highly efficient methods\footnote{For a complete review about currently available technologies for QFT calculations, see Ref. \cite{Heinrich:2020ybq} and references therein.}. Special emphasis was recently put in the development of four dimensional methods \cite{Gnendiger:2017pys,Heinrich:2020ybq,TorresBobadilla:2020ekr}, with the purpose of achieving a seamless combination of algebraic/analytic strategies and numerical integration directly in the four physical dimensions of the space-time. Overcoming the current precision frontier will certainly require to push these methods to their limits.

Future collider experiments require challenging theoretical predictions from full calculations at order $n$ in the perturbative expansion (N$^n$LO), with $n\geq 3$.
One foreseeable complication is directly related to the appearance of several scales in multiloop multileg Feynman integrals. These objects are analytically known for specific processes and kinematic configurations up to two loops with a limited number of scales and very few examples of three-loop amplitudes are starting to pop-up \cite{Caola:2020dfu,Caola:2021rqz}. At this point, it is very unlikely that all the required ingredients will become analytically available and the development of novel numerical approaches seems unavoidable.

Given present and future needs, a major progress in solving (at least) two challenges is necessary. On one side, new computational techniques must be developed, exploiting the fundamental properties of QFT and the mathematical concepts behind scattering amplitudes. On the other side, these new methods must be implemented in efficient event generators, capable of overcoming the bottlenecks of the currently available hardware. 

Regarding the first challenge, we focus on the Loop-Tree Duality (LTD) formalism~\cite{Catani:2008xa,Rodrigo:2008fp,Bierenbaum:2010cy,Bierenbaum:2012th,Buchta:2014dfa,Buchta:2015wna,Buchta:2015xda,Jurado:2017xut,Tomboulis:2017rvd,Plenter:2019jyj,Plenter:2020lop,Plenter:2022zxk}, which exhibits very attractive mathematical properties and a manifestly causal physical interpretation. Multiloop scattering amplitudes are transformed in LTD into the so-called \emph{dual amplitudes} by integrating out one component of each of the loop momenta through the Cauchy's residue theorem. In this way,
physical observables are expressed in terms of Euclidean integrals that combine loop and tree-level contributions
(as well as renormalization counter-terms) to achieve a fully local cancellation of singularities~\cite{Hernandez-Pinto:2015ysa,Sborlini:2016gbr,Sborlini:2016hat,Driencourt-Mangin:2019aix,Driencourt-Mangin:2019sfl,Driencourt-Mangin:2019yhu,Prisco:2020kyb}. Besides the possibility of a local regularization, including the simultaneous cancellation of infrared and ultraviolet singularities, the LTD formalism fully exploits causality in QFT. A recent reformulation~\cite{Aguilera-Verdugo:2019kbz,Verdugo:2020kzh,Aguilera-Verdugo:2020kzc,Ramirez-Uribe:2020hes,Aguilera-Verdugo:2020nrp,TorresBobadilla:2021ivx,Sborlini:2021owe,TorresBobadilla:2021dkq,Capatti:2019ypt,Capatti:2019edf,Runkel:2019yrs,Runkel:2019zbm,Kromin:2022txz} showed that multiloop scattering amplitudes and Feynman integrals can be represented in terms of a subset of cut diagrams that generalize the well-known Cutkosky's rules \cite{Cutkosky:1960sp}. In particular, a manifestly causal representation can be directly obtained by decomposing the original Feynman diagrams into binary connected partitions in the equivalence class of topologies defined by collapsing propagators into edges (or multi-edges) \cite{Sborlini:2021owe,Sborlini:2021nqu}. As a result, a well defined geometrical algorithm is available and a strong connection between causality and directed acyclic graphs can be established~\cite{Ramirez-Uribe:2021ubp}.

The other challenge is related to an efficient calculation of Feynman integrals, overcoming current hardware limitations. In this direction, the development of novel strategies for classically hard problems based on quantum algorithms (QAs) is gaining momentum across different areas. For instance, there are several ideas that exploit the potential speed-up of quantum computers, such as database querying through Grover's algorithm \cite{Grover:1997fa}, the famous Shor's algorithm for factorization of large integers~\cite{Shor:1994jg} or Hamiltonian minimization through quantum annealing \cite{PhysRevE.58.5355}. In the context of particle physics, QAs are often applied to solve problems related to lattice gauge theories~\cite{Jordan:2011ne,Banuls:2019bmf,Zohar:2015hwa,Byrnes:2005qx,Ferguson:2020qyf,Kan:2021nyu}. Recent applications for high-energy colliders include jet identification and clustering~\cite{Wei:2019rqy,Pires:2021fka,Pires:2020urc,deLejarza:2022bwc,Delgado:2022snu,Barata:2022wim,Barata:2021yri}, determination of parton densities (PDFs)~\cite{Perez-Salinas:2020nem}, simulation of parton showers~\cite{Williams:2021aji,Gustafson:2022xwt}, 
anomaly detection~\cite{Ngairangbam:2021yma}, and integration of elementary particle processes~\cite{Agliardi:2022ghn}. This list is rapidly growing, since QAs are suitable for several uses, especially those involving minimization problems.

With this panorama in mind, the purpose of this article is to explore the application of QAs for unveiling the causal structure of multiloop scattering amplitudes and Feynman integrals, or equivalently acyclic configurations
of directed graphs. Our strategy consists in exploiting the properties of the adjacency matrix in graph theory to build a Hamiltonian which weights the cost of different momentum flow configurations. Explicitly, causal configurations are those with minimum energy, so we select them by identifying the minima of the Hamiltonian. This identification is implemented through a Variational Quantum Eigensolver (VQE) \cite{Peruzzo_2014,McClean_2016,Tilly:2021jem}, namely a hybrid quantum-classical algorithm that seeks the ground state of a given operator.

The classical problem of identifying or counting directed acyclic graphs is \emph{\#P-hard}, see Ref. \cite{doi:10.1137/0607036}. Explicitly, given a graph $G=(V,E)$ with $V$ vertices and $E$ edges, there are $2^{|E|}$ possible directed graphs that must be checked for cycles.
Efficient classical algorithms have been proposed in the literature: e.g., Ref.~\cite{OLIVEIRA2019655} gives an example of a Fixed Parameter Tractable algorithm, whose scaling depends on a single structural parameter of the graph, see also references therein. This algorithm consists of a Binary Decision Diagram algorithm~\cite{bryant1986graph}, which performs asymptotically in ${\cal O}(2^{p_w^2/4})$ running time per solution, with the \emph{path-width} $p_w$ always smaller than the input size of the graph, becoming closer to it for dense graphs (i.e., maximally connected graphs). Instead, the motivation of our paper is to encode the complexity of the graph by means of a Hamiltonian, i.e. a cost function to be minimized in order to find the directed acyclic configurations. Since VQE is expected to offer a speed-up in minimization problems compared to purely classical algorithms, we explore up to what extent this speed-up remains in the detection of directed acyclic graphs.

The outline of this paper is the following. In Sec. \ref{sec:Causality}, we present a brief introduction to LTD, and we offer a description of its manifestly causal representation. Then, in Sec. \ref{sec:Causalflow}, we discuss the geometrical aspects of the causal configurations, presenting explicit examples in App. \ref{app:CAUSALEJEMPLO}. After that, we introduce the Hamiltonian approach in Sec. \ref{sec:Hamiltonian}, putting special emphasis in the connection with the adjacency matrix of directed acyclic graphs. We offer a definition of the \emph{loop Hamiltonian} and its classical reconstruction algorithm in Secs. \ref{ssec:loopHamiltonian} and \ref{ssec:Reconstruction}, respectively. We discuss subtleties of the encoding of vertex registers in App. \ref{app:Encoding}. Then, we discuss the implementation of the VQE in Sec. \ref{sec:NumericalHamiltonian}, presenting an explicit example for a representative two-loop topology in Sec. \ref{ssec:2eloopHam}. The full list of Hamiltonians for the topologies studied in this article is given in App. \ref{app:Hamiltonians}. Right after in Sec. \ref{ssec:IterativeVQE}, we discuss an improved strategy based on multiple runs of the VQE to collect, step by step, all the possible causal solutions. In Sec. \ref{sec:ConnectionGH}, we carefully compare the VQE approach with the Grover's based algorithm described in Ref.~\cite{Ramirez-Uribe:2021ubp}, paying attention to the resources required for a successful implementation in (real) quantum devices. Finally, conclusions and further research directions are presented in Sec. \ref{sec:Conclusions}.

%%%%%%%%%%%%%%%%%%%%%%%%%%%%%%%%%%%%%%%%%%%%%%%%%%%%%%%%%%%%%%%%%%%%%%%%%%%%%%
%%%%%%%%%%%%%%%%%%%%%%%%%%%%%%%%%%%%%%%%%%%%%%%%%%%%%%%%%%%%%%%%%%%%%%%%%%%%%%
\section{Loop-Tree Duality and Causality}
\label{sec:Causality}
To reach highly precise theoretical predictions, it is necessary to deal with multiloop scattering amplitudes and the corresponding Feynman integrals. In the Feynman representation, the most general $L$-loop scattering amplitude with $P$ external particles is given by
\beq
{\cal A}_F^{(L)} = \int_{\ell_1 \ldots \ell_L} {\cal N} 
\big(\{\ell_s\}_L, \{p_j\}_P\big) \prod_{i=1}^n G_F(q_i)~,
\label{eq:FeynmanRep}
\eeq
where $q_i$ with $i\in \{1,\ldots,n\}$ are the momenta flowing through each Feynman propagator, $G_F(q_i)=(q_i^2-m_i^2+ \imath 0)^{-1}$, and ${\cal N}$ represents a numerator, whose specific form depends on the topologies of the different diagrams, the interaction vertices and the nature of the particles that propagate inside the loops.
Regarding the momenta of the internal particles, they are linear combinations of the primitive loop momenta associated to the integration variables in the loop, $\ell_s$ with $s\in \{1,\ldots,L\}$), and the external momenta, $p_j$ with $j\in\{1,\ldots, P\}$.
Given a Feynman diagram, we can group the different internal momenta into \emph{sets}: two lines are said to belong to the same set if their propagators involve the same linear combination of primitive loop momenta~\cite{Verdugo:2020kzh}. This is useful for achieving an efficient classification of diagrams according to their causal structure~\cite{Aguilera-Verdugo:2020kzc,Ramirez-Uribe:2020hes,Aguilera-Verdugo:2020nrp}.

The denominator of \Eq{eq:FeynmanRep} characterizes the singular structure of the scattering amplitude, and singularities provide important information to simplify loop calculations. In this direction, the Loop-Tree Duality (LTD) \cite{Catani:2008xa,Rodrigo:2008fp} makes use of Cauchy's residue theorem (CRT) to reduce the dimensionality of the integration domain, enabling the possibility of transforming it into an Euclidean space. By means of an iterated application of CRT, we can get rid of one integration variable per loop: this is equivalent to say that the LTD representation of a $L$-loop amplitude is obtained by cutting (or setting on-shell) $L$ internal lines. The effect of cutting a line is the modification of the infinitesimal complex prescription, leading to the so-called \emph{dual propagators} \cite{Catani:2008xa,Rodrigo:2008fp}. It is important to highlight that this modified prescription plays a crucial role to preserve causality, as we will discuss later.

%%%%%%%%%%%%%%%%%%%%%%%%%%%%%%%%%%%%%%%%%%%%%%%%%%%%%%%%%%%%%%%%%%%%%%%%%%%%%%
%%%%%%%%%%%%%%%%%%%%%%%%%%%%%%%%%%%%%%%%%%%%%%%%%%%%%%%%%%%%%%%%%%%%%%%%%%%%%%

The traditional formulation of LTD leads to the so-called \emph{dual representation}, in which there is one contribution for each possible connected  tree obtained by setting on-shell $L$ internal propagators. If we study the structure of the denominators of each dual term, we immediately realize the presence of divergences that do not correspond to any physical configuration. These unphysical singularities are spurious, and they vanish when we add all the dual contributions together. This is the so called manifestly causal representation within LTD~\cite{Verdugo:2020kzh,MANIFESTLYCAUSAL}.

Before moving on, let us recall one crucial fact about the Feynman propagators. If we define the positive on-shell energy
\beq
\qon{i} = \sqrt{\vec{q_i}^2+m_i^2- \imath 0} \, ,
\eeq
then the propagator can be written as
\beq
G_F(q_i)= \frac{1}{q_{i,0}-\qon{i}} \times \frac{1}{q_{i,0}+\qon{i}} \, .
\label{eq:GFdecomposed}
\eeq
This decomposition suggests that each Feynman propagator encodes a quantum superposition of two on-shell modes, with positive and negative energy respectively. The so-called \emph{causal representations} are obtained by consistently aligning these modes. Furthermore, this superposition also motivates the identification of each internal propagator with a qubit, as we will explain later.

As carefully discussed in Refs.~\cite{Verdugo:2020kzh,Aguilera-Verdugo:2021nrn}, the calculation of the nested residues within the LTD formalism leads to a manifestly causal representation of multiloop multileg scattering amplitudes. Thus, it can be shown that \Eq{eq:FeynmanRep} is equivalent to
\beq
{\cal A}_D^{(L)} = \int_{\vec \ell_1 \ldots \vec \ell_L} 
\frac{1}{x_n} \sum_{\sigma  \in \Sigma} {\cal N}_{\sigma}\, \prod_{i=1}^{n-L} \, \frac{1}{\lambda_{\sigma(i)}^{h_{\sigma(i)}}} \
+ (\lambda^+_p \leftrightarrow \lambda^-_p)~,
\label{eq:CausalRepresentation}
\eeq
with $x_n = \prod_n 2\qon{i}$, $h_{\sigma(i)} = \pm 1$, and
\beq
\int_{\vec \ell_s} = -\mu^{4-d} \int  \frac{d^{d-1} \ell_s}{(2\pi)^{d-1}} \, ,
\label{eq:MeasureEuclidean}
\eeq
the integration measure in the loop three-momentum space. It turns out that Eq. (\ref{eq:CausalRepresentation}) only involves denominators with on-shell energies, added together in same-sign combinations within the so-called \emph{causal propagators}, 
$1/\lambda_{\sigma(i)}^{h_{\sigma(i)}}$, with
\beq
\lambda_{\sigma(i)}^{h_{\sigma(i)}} \equiv \lambda_p^\pm = \sum_{i\in p} \qon{i} \pm k_{p,0}~,
\eeq
where $\sigma(i)$ includes information about the partition $p$ of the set of on-shell energies and the corresponding orientation of the energy components of the external momenta, $k_{p,0}$. Each causal propagator is in a one-to-one correspondence with any possible threshold singularity of the amplitude ${\cal A}_F^{(L)}$, which contains overlapped thresholds. These are known as \emph{entangled causal thresholds}: $\sigma(i)$ indicates the set of causal propagators that can be simultaneously entangled. The set $\Sigma$ in \Eq{eq:CausalRepresentation} contains all the combinations of allowed causal entangled thresholds, which involves overlapped cuts with aligned momenta flow. 

An important advantage of the representation shown in \Eq{eq:CausalRepresentation} is the absence of spurious nonphysical singularities. In fact, since causal propagators involve same-sign combinations of positive on-shell energies, the only possible singularities are related to IR/UV and physical thresholds. More details about the manifestly causal LTD representation and its benefits can be found in Refs.~\cite{Aguilera-Verdugo:2020kzc,Ramirez-Uribe:2020hes,TorresBobadilla:2021ivx,Sborlini:2021owe}.

%%%%%%%%%%%%%%%%%%%%%%%%%%%%%%%%%%%%%%%%%%%%%%%%%%%%%%%%%%%%%%%%%%%%%%%%%
%%%%%%%%%%%%%%%%%%%%%%%%%%%%%%%%%%%%%%%%%%%%%%%%%%%%%%%%%%%%%%%%%%%%%%%%%
\section{Geometric interpretation of causal flows}
\label{sec:Causalflow}
In order to identify the causal configurations of a multiloop Feynman diagram in a quantum device, it is necessary to translate the problem into a suitable language. The geometrical formulation of the manifestly causal LTD representations~\cite{Sborlini:2021owe} provides a clear and intuitive interpretation. In the following, we briefly describe the most relevant features of this formalism, recalling some basic concepts and definitions as presented in Refs.~\cite{Ramirez-Uribe:2021ubp,Sborlini:2021owe,Sborlini:2021nqu}.

Any scattering amplitude can be represented starting from Feynman diagrams built from interaction vertices and internal lines (or propagators) connecting those vertices. Regarding causality, those lines that connect the same vertices can be merged into a single \emph{edge}, leading to \emph{reduced Feynman graphs} because the only allowed configurations which are causal are those in which all the momentum flows of these propagators are aligned in the same direction. Then, reduced graphs are built from vertices and \emph{edges}, as described in Refs.~\cite{TorresBobadilla:2021ivx,Sborlini:2021owe,TorresBobadilla:2021dkq}, and their causal structure turns out to be equivalent to that of the dual representation of the original Feynman diagrams~\cite{Aguilera-Verdugo:2020nrp}. The number of loop integration variables in the LTD representation given by \Eq{eq:CausalRepresentation} is directly related to the topological independent loops present at the level of the Feynman diagram. However, the reduced Feynman graph has a fewer number of graphical loops, also called \emph{eloops}, since bunches of propagators connecting the same vertices were collapsed to a single edge. Furthermore, it turns out that vertices, edges and eloops are the only required ingredients to obtain the causal representation of any multiloop scattering amplitude \cite{Sborlini:2021owe,TorresBobadilla:2021ivx}.

Specifically, given a reduced Feynman graph, the associated causal propagators $1/\lambda_p^{\pm}$ correspond to the set of connected binary partitions of vertices. Also, they can be graphically interpreted as lines cutting the diagrams into two disconnected pieces, in a full analogy with Cutkosky's formulation \cite{Cutkosky:1960sp}. Then, the representation in Eq. (\ref{eq:CausalRepresentation}) can be constructed from all the possible causal compatible combinations of $k$ causal propagators, the so-called \emph{causal entangled thresholds}. The number $k$ is known as the order of the diagram and defined as $k=V-1$ \cite{Verdugo:2020kzh,TorresBobadilla:2021ivx,Sborlini:2021owe}. It naturally induces a topological classification of families of Feynman diagrams, or equivalently, reduced Feynman graphs. The family with $k=1$ is known as Maximal Loop Topology (MLT) \cite{Verdugo:2020kzh} and only involves two vertices; a Next-to-Maximal Loop Topology (NMLT) has three vertices, and so on.

It was shown that causal entangled thresholds can be identified by imposing geometrical selection rules. These rules are deeply connected to the algebraic formulation presented in Refs.~\cite{TorresBobadilla:2021ivx,Benincasa:2021qcb}, and they establish that:
\begin{enumerate}
 \item When considering all the $k$ causal thresholds that can be simultaneously entangled, all the edges can be set on shell simultaneously. This is equivalent to impose that the causal entangled thresholds depends on the on-shell energies $\qon{i}$ of all the edges.
 \item Two different causal propagator $\lambda_i$ and $\lambda_j$ can be simultaneously entangled if they do not cross each other. In other words, this means that the associated partition of vertices do not intersect or are totally included one in the other.
 \item The momentum of the edges that crosses a given binary partition of vertices must be consistently aligned, i.e., momentum must flow from one partition to a different one.
\end{enumerate}
A careful explanation of these causal rules and their geometric interpretation was presented in Ref.~\cite{Sborlini:2021owe}, including pedagogical examples and practical cases of use. By imposing these three rules on the set of all the possible combinations of $k$ thresholds, we construct $\Sigma$, namely the set of all causal entangled thresholds leading to \Eq{eq:CausalRepresentation}. Furthermore, as previously shown in Ref.~\cite{Ramirez-Uribe:2021ubp}, the third condition is equivalent to ordering the internal edges in such a way that the reduced Feynman diagram is a \emph{directed acyclic graph}. This means that there cannot be closed cycles, allowing the information to propagate consistently from one partition of the diagram to the other; this condition is necessary for having a causal-compatible partition. For this reason, given a reduced Feynman graph, it is crucial to efficiently detect all the associated \emph{directed acyclic graphs}, since they are a vital ingredient to reconstruct the LTD causal representation. To clarify this discussion, we present an explicit application example in App. \ref{app:CAUSALEJEMPLO}.

%%%%%%%%%%%%%%%%%%%%%%%%%%%%%%%%%%%%%%%%%%%%%%%%%%%%%%%%%%%%%%%%%%%%%%%%%%%%%%
%%%%%%%%%%%%%%%%%%%%%%%%%%%%%%%%%%%%%%%%%%%%%%%%%%%%%%%%%%%%%%%%%%%%%%%%%%%%%%
\section{Hamiltonian formalism for causal-flow identification}
\label{sec:Hamiltonian}
The first proof-of-concept of a quantum algorithm for Feynman loop integrals was presented in Ref.~\cite{Ramirez-Uribe:2021ubp}, managing to successfully unfold the causal configurations of multiloop Feynman diagrams. 
The implementation follows a modified Grover's quantum algorithm to identify the presence of directed acyclic configurations, representing the causal solutions, over different subloops of the multiloop topologies.

The aim of this Section is to define a Hamiltonian whose ground state corresponds to a superposition of all the acyclic configurations of a given multiloop Feynman diagram\footnote{To avoid any possible confusion, we would like to emphasize that this is not the Hamiltonian of the theory from which the Feynman rules are extracted, but rather a function whose minimization sets by construction the causal orientations of the diagram.}. We first discuss the construction of such Hamiltonian based on direct inspection of the multiloop topology, and then we discuss a more general approach based on the \emph{adjacency matrix} of a graph, an object that concentrates all the information regarding the orientation of edges and the vertices that they connect within a reduced Feynman graph (as defined in Sec. \ref{sec:Causalflow}). We also give a more formal presentation, relying on concepts and notations from graph theory.

%%%%%%%%%%%%%%%%%%%%%%%%%%%%%%%%%%%%%%%%%%%%%%%%%%%%%%%%%%%%%%%%%%%%%%%%%%%%%%
%%%%%%%%%%%%%%%%%%%%%%%%%%%%%%%%%%%%%%%%%%%%%%%%%%%%%%%%%%%%%%%%%%%%%%%%%%%%%%
\subsection{Cycle detection via Hamiltonian optimization}
\label{ssec:loopHamiltonian}
Let us consider a generic undirected graph $G=(V,E)$, 
with $V$ a set of distinct vertices and $E\subset V \times V$ a set of edges (also known as \emph{links}). Given the graph $G$, one can build a set $\mathcal{D}_G$ of $2^{|E|}$ possible directed graphs 
where we associate a specific direction to each link. We will denote the subset of directed acyclic graphs as $\widetilde{\mathcal{D}}_G \subset \mathcal{D}_G$.

In order to formulate this problem as a Hamiltonian optimization,
we first fix the encoding for all the possible (classical) solutions 
as elements of the computational basis as follows. We start by establishing a conventional orientation to each edge, 
selecting a specific element from the set of directed graphs $G_0 \in \mathcal{D}_G$. This set is associated to the state $\ket{00\dots0}\in \mathcal{H}_E$, 
mapping each qubit to an edge as done in  Ref.~\cite{Ramirez-Uribe:2021ubp}: the single qubit state $\ket{0}_e$ for the edge $e\in E$
corresponds to a link with the same orientation of $e$ in the conventional graph $G_0$,
while the state $\ket{1}_e$ would correspond to the opposite orientation. The problem can then be cast in the search of 
all the possible states of the computational basis of edge orientations associated 
to directed acyclic graphs $\widetilde{\mathcal{D}}_G$, 
whose span generates a Hilbert space $\widetilde{\mathcal{H}}_E\subset \mathcal{H}_E$.

After these definitions, we need to identify the set $\Gamma_{G_0}$ of all the possible oriented cycles. In this notation, the subscript $G_0$ means that we will use the reference graph to specify the orientations of those cycles. Here, \emph{cycle} refers to \emph{simple cycles} (i.e. a closed loop in which a vertex appears only once). If the set $\Gamma_{G_0}$ of directed loops is already known, one can build the so-called \emph{loop Hamiltonian}, which is a Hamiltonian on the space $\mathcal{H}_E$, diagonal in the computational basis described above. We proceed by penalizing oriented cycles using diagonal projectors, i.e.
\begin{equation}\label{eq:loopHam_hardcoded}
    H_G = \sum_{\gamma\in\Gamma_{G_0}} \prod_{e\in \gamma} \pi^{s(e;G_0)}_e~,
\end{equation}
where $s(e;G_0)$ evaluates to $0$ or $1$ according to the orientation of $e\in \gamma$
relative to the conventional graph orientation $G_0$, and the $\pi$ symbol 
identifies the projector operators on single edge registers,
\begin{equation}
\begin{aligned} 
    \pi^0_e &\equiv \ketbra{0}{0}_e=\frac{1}{2} {(I+Z)}_e~,\\
    \pi^1_e &\equiv \ketbra{1}{1}_e=\frac{1}{2}{(I-Z)}_e~,
\end{aligned} 
\label{eq:Projector1}
\end{equation}
with $Z$ the Pauli matrix in the $z$-axis.

%++++++++++++++++++++++++++++++++
\begin{figure}[t]
\centering
\includegraphics[scale=1]{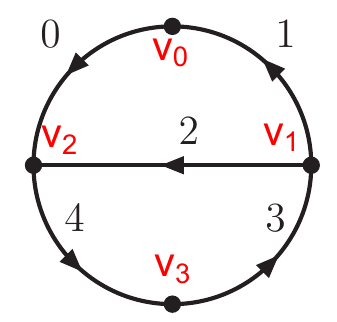}
\caption{Example of a diagram with two eloops and four vertices representing external particles (Topology A). Edges are directly numbered, and vertices are explicitly indicated as $v_i$. 
\label{fig:TwoLoopsQC}}
\end{figure}
%++++++++++++++++++++++++++++++++

To provide an explicit example, let us consider Fig.~\ref{fig:TwoLoopsQC}. It depicts a graph $G_0$ (i.e., $G$ equipped with a conventional orientation), where one can identify by direct inspection the set of all the possible directed loops:
\begin{align}
    \Gamma_{G_0} = 
    \Big\{ &\big[\bar 0,2,\bar 1\big], \big[0,\bar 2,1\big], \big[4,3,2\big], [\bar 4,\bar 3,\bar 2\big], \nn\\
    & \big[1,4,3,0\big],\big[\bar 1,\bar 4,\bar 3,\bar 0\big] \Big\}~,
\end{align}
where the presence or absence of the bar identifies the orientation (i.e., $s(e,G_0)=0$, $s(\bar e,G_0)=1$). Then, the loop Hamiltonian has the following form:
\begin{align}\label{eq:HamiltonianFIG1}
    H_G &= \pi^1_0 \pi^0_2 \pi^1_1 + \pi^0_0 \pi^1_2 \pi^0_1 + \pi^0_4 \pi^0_3 \pi^0_2 \nn\\
    &+ \pi^1_4 \pi^1_3 \pi^1_2 + \pi^0_1 \pi^0_4 \pi^0_3\pi^0_0 +\pi^1_1 \pi^1_4 \pi^1_3\pi^1_0.
\end{align}
Notice that a term in $H_G$ acts trivially on a non-indicated edge, i.e.
\beq
\pi^1_0 \pi^0_2 \pi^1_1 \equiv I_4 \otimes I_3 \otimes \pi^0_2 \otimes \pi^1_1 \otimes \pi^1_0  \, ,
\eeq 
with $I_i$ corresponding to the identity operator on ${\cal H}_i$. This loop Hamiltonian will evaluate to $0$ for any solution in $\widetilde{\mathcal{D}}_G$ of graphs without oriented cycles (i.e., $\Ker(H_G) = \widetilde{\mathcal{H}}_E$), while the other directed graphs are associated to a strictly greater than $0$ integer eigenvalue. Moreover, the same oriented graph could have more than one closed cycle, and this Hamiltonian counts the number of these oriented cycles.

%%%%%%%%%%%%%%%%%%%%%%%%%%%%%%%%%%%%%%%%%%%%%%%%%%%%%%%%%%%%%%%%%%%%%%%%%%%%%%
%%%%%%%%%%%%%%%%%%%%%%%%%%%%%%%%%%%%%%%%%%%%%%%%%%%%%%%%%%%%%%%%%%%%%%%%%%%%%%
\subsection{Classical algorithm for reconstructing the loop Hamiltonian}
\label{ssec:Reconstruction}
The adjacency matrix of a graph provides a powerful tool to further characterise the associated multiloop topology, and it can be used when the loop set is not straightforward to compute. Classically, the adjacency matrix $A = || a ||$ of a directed graph of $N=|V|$ vertices (labelled in an arbitrary manner) is defined such that $a_{i_1 i_2} = 1$ ($i_1, i_2 = 1, \ldots, N$) if there is a directed edge from $i_1$ to $i_2$, and $0$ otherwise. Note that a cycle of length $ \ell $ along the vertices $ i_0, \ldots, i_{\ell-1}, i_\ell = i_0 $ implies $ a_{i_0 i_1} \times \ldots \times a_{i_{\ell-1} i_\ell} = 1 $.  
Moreover, if $ A \in \mathbb{Z}_2^{N \times N}$, the following properties are equivalent:
\begin{enumerate}
 \item the graph of which $ A $ is the adjacency matrix is acyclic;
 \item $A^N = 0$, i.e. $A$ is nilpotent (which is stronger than $\text{tr} (A^N) = 0$);
 \item $A$ is equivalent to a triangular matrix, where \emph{equivalence} has a precise algebraic meaning, corresponding to the existence of a relabelling;
 \item the eigenvalues of $A$ are all zero, which directly implies $ \text{tr} (A^k) = 0 $ for any positive integer $k$.
\end{enumerate}
We can exploit these equivalent properties to build suitable candidates for Hamiltonians. For this purpose, let us consider a graph $G=(V,E)$ and a corresponding conventionally oriented graph $G_0=(V,\bar{E})$. Then, we define an extended Hilbert space of vertices and edges $\mathcal{H}_{VE} \equiv \mathcal{H}_V \times \mathcal{H}_E$, and we build a linear operator acting on it and encoding the information about the oriented adjacency matrix:
\begin{align}
    A \equiv \sum_{e\equiv(v_0,v_1)\in \bar{E}} \Big[\sigma^-_{v_0} \pi^0_{e} \sigma^+_{v_1} + \sigma^-_{v_1} \pi^1_{e} \sigma^+_{v_0} \Big] \, ,
\end{align}
where $\sigma^\pm = (X\pm i\, Y)/2$ are the \emph{ladder operators}. This construction can be interpreted as a sum of hopping terms from the tail to the head vertices of each oriented edge $e$, with the edge projector operators $\pi^s_e$ keeping track of the orientation relative to the conventional graph. Thus, joining two adjacent edges would give nontrivial contributions only if the orientation is the same for both edges 
(i.e., if the head of one is joined with the tail of the other), and that is ensured by the properties of the ladder operators on the corresponding vertex register. Therefore, the $n$-th power of the $A$ operator will have non-zero diagonal terms
in the vertex side only if all the ladder operators are completely \emph{contracted} and only 
projectors $\pi^s$ appear. On the edge side, these terms identify oriented cycles (or disjoint products of oriented cycles), so that a possible loop Hamiltonian,
whose kernel is the span of directed acyclic graph states $\widetilde{\mathcal{H}}_E$, 
can be written as follows:
\begin{equation}\label{eq:loopHam_trn}
    H_G = \sum\limits_{n=1}^{M_G} \text{tr} (A^n) \, ,
\end{equation}
where the trace runs only over the vertex space, and $M_G$ is the maximal length of loops in $G$, i.e., $ M_G = |V| $. An oriented cycle going across $j$ vertices corresponds to a term in this sum when $n=j$. The loop Hamiltonian built in this way has the same terms as the loop Hamiltonian in Eq.~\eqref{eq:loopHam_hardcoded}, plus some terms representing disjoint products of oriented cycles (e.g., two oriented cycles going respectively across $j_1$ and $j_2$ vertices and corresponding to a term in the sum above when $n=j_1+j_2$). Also, it includes coefficients in front of the different products of projectors $\pi$, which are related to topological properties of the graph under consideration (see, e.g., \Eq{eq:HamiltonianoEUnrestricted} below). In any case, the ground state will be associated with the subspace of directed acyclic graphs. For the practical implementation, we will rely on the normalized version of $H_G$ where all the coefficients in front of the $\pi$ operators are exactly 1.

It is worth mentioning that, to derive a loop Hamiltonian, in principle any positive-definite function of $A$ can be used. For example,
\begin{equation}\label{eq:loopHam_treA}
H^\prime_G = \text{tr} (e^A-I) \, ,
\end{equation}
would have a different spectrum, but exactly the same kernel as the loop Hamiltonian in Eq.~\eqref{eq:loopHam_trn}. Given the relation ${\rm tr} ( A^j ) = \sum^N_{i=1} \lambda^j_i$, where $\lambda_i$ ($i=1, \ldots, N$) are the eigenvalues of the adjacency matrix, one sees that cycles are deeply connected to spectral properties of graphs. So, the problem we have at hand consists in knowing whether any of these eigenvalues is non-zero, in which case cycles would be present. Since the Hamiltonians defined above are related to the eigenvalues, they are independent of the specific labelling of the graph in consideration.

To conclude this section, note that a discussion about efficient encoding methods for the vertex space is presented in App.~\ref{app:Encoding}, having in mind the possibility of a Hamiltonian reconstruction by means of a purely quantum algorithm. Also, we would like to mention that the number of terms in the loop Hamiltonian does not depend on the number of external legs. As shown in App. \ref{app:Hamiltonians}, more complex topologies require more terms because each term of the Hamiltonian penalizes the presence of cycles. Thus, more loops leads to more possible cycles to be checked, and this translates into more terms in the Hamiltonian, but more external legs do not generate additional cycles. Computing the Hamiltonian in a fully quantum way, as we discuss in App.~\ref{app:Encoding}, could allow to speed-up the codification of the characteristics of the graph for the posterior minimization with a quantum device.

%%%%%%%%%%%%%%%%%%%%%%%%%%%%%%%%%%%%%%%%%%%%%%%%%%%%%%%%%%%%%%%%%%%%%%%%%%%%%%
%%%%%%%%%%%%%%%%%%%%%%%%%%%%%%%%%%%%%%%%%%%%%%%%%%%%%%%%%%%%%%%%%%%%%%%%%%%%%%
\section{VQE implementation and numerical results}
\label{sec:NumericalHamiltonian}
We now discuss the application of the Variational Quantum Eigensolver (VQE) algorithm~\cite{Peruzzo_2014,McClean_2016} to find directed acyclic graphs given a fixed topology. Fixing a positive orientation for an 
arbitrary link (the first one for example), 
the $|E|-1$ remaining qubits 
encode the direction of the other links, 
i.e., the states considered have the 
form $\ket{\psi} \otimes \ket{0}$ with the 
\emph{restricted} loop Hamiltonian $H_{E/\{e_0\}} \equiv H_E \rvert_{s(e_0,G_0) = 0}$ acting only on 
$\ket{\psi}$. This is possible without loss of 
generality, since the solutions to the unrestricted
problem can be found by adding a copy of the 
solutions to the restricted problem $\{\ket{\psi_j}\} \subset \widetilde{\mathcal{H}}_{E/\{e_0\}} \cap \mathbb{Z}_2^{|E|-1}$  
with all links flipped:
\beq
\big\{\ket{\psi_j} \otimes \ket{0} \big\} \cup \big\{\big( X^{\otimes {(|E|-1)}}\ket{\psi_j} \big) \otimes \ket{1} \big\}
\subset \widetilde{\mathcal{H}}_{E} \cap \mathbb{Z}_2^{|E|} \, ,
\eeq
where $\big( X^{\otimes {(|E|-1)}}\ket{\psi_j} \big)$ corresponds to applying a tensor product of \texttt{NOT} gates to the first $|E|-1$ bits to invert the directions of the links/edges described in $\ket{\psi_j}$.

To explain how the proposed algorithm works, let us consider again the topology shown in 
Fig.~\ref{fig:TwoLoopsQC}, where $|E|=5$. The unrestricted Hamiltonian obtained from Eq. (\ref{eq:loopHam_trn}) is
\begin{align}\label{eq:HamiltonianoEUnrestricted}
    H_E &= 3\pi_{0}^1 \, \pi_{1}^1 \, \pi_{2}^0 \, 
    + 3\pi_{0}^0 \, \pi_{1}^0 \, \pi_{2}^1 \, 
    + 4\pi_{0}^0 \, \pi_{1}^0 \, \pi_{3}^0 \, \pi_{4}^0 \,  \nn\\[3pt]
    &+ 3\pi_{2}^0 \, \pi_{3}^0 \, \pi_{4}^0 \, 
    + 4\pi_{0}^1 \, \pi_{1}^1 \, \pi_{3}^1 \, \pi_{4}^1 \, 
    + 3\pi_{2}^1 \, \pi_{3}^1 \, \pi_{4}^1 \, \, \, ,
\end{align}
where we point out the presence of nontrivial coefficients in front of each term, that results from having identical terms in Eq. (\ref{eq:loopHam_trn}). As explained before, the ground state will have always 0 energy and it will be associated to the set of directed acyclic graphs, independently of the spectrum of $H_E$. For the sake of simplicity, from now on, we will work with the normalized Hamiltonian, eliminating the positive coefficients seen in \Eq{eq:HamiltonianoEUnrestricted}.

Then, we can fix the orientation of the first qubit in agreement with the initial choice which implies
\beq
\pi_0^0 \ket{\psi} = \ket{\psi} \, , \quad \pi_0^1 \ket{\psi} = 0 \, ,
\eeq
with $\ket{\psi} \in \{\ket{\psi'} \otimes \ket{0}\}$. This is equivalent to set $\pi_0^0 \equiv I_0$ and remove all the terms proportional to $\pi_0^1$. Explicitly
\begin{align}
    H_{E/\{e_0\}} 
    &= \pi_{1}^0 \, \pi_{2}^1 \, 
    + \pi_{1}^0 \, \pi_{3}^0 \, \pi_{4}^0 \, \nn\\
    &+ \pi_{2}^0 \, \pi_{3}^0 \, \pi_{4}^0 \, 
    + \pi_{2}^1 \, \pi_{3}^1 \, \pi_{4}^1 \, ,
\label{eq:HamiltonianoERestricted}
\end{align}
which acts over the reduced edge space. The computational basis of the restricted problem represents the $2^{|E|-1} = 16$ possible directed graphs, of which only $9$ are acyclic 
(i.e., the solutions to our minimization problem). In practice, all the $9$ solutions of the reduced problem can be found by applying the VQE where we minimize $\expval{H_{E/\{e_0\}}}$ with any of the Hamiltonians built as discussed in the previous sections. As argued above, the space of solutions corresponds to the intersection between the canonical basis and the eigenspace associated with the eigenvalue $0$. A single run of the VQE converging to $\expval{H_{E/\{e_0\}}}=0$ (within precision), will, in general, result in a superposition of states of the computational basis in $\Ker H_{E/\{e_0\}}$. In analogy to the Grover's approach, we collect solutions
whose probability appears above a certain threshold $\lambda$.
However, in this case, there is no guarantee that all possible solutions are found in the superposition of a single run. In the following, we will explain a possible strategy to overcome this difficulty, based on running the VQE multiple times.

%++++++++++++++++++++++++++++++++++++++++++++++++++++++++++++++++++++++++++++++++++++
\begin{figure*}
    \centering
    \includegraphics[width=0.47\linewidth]{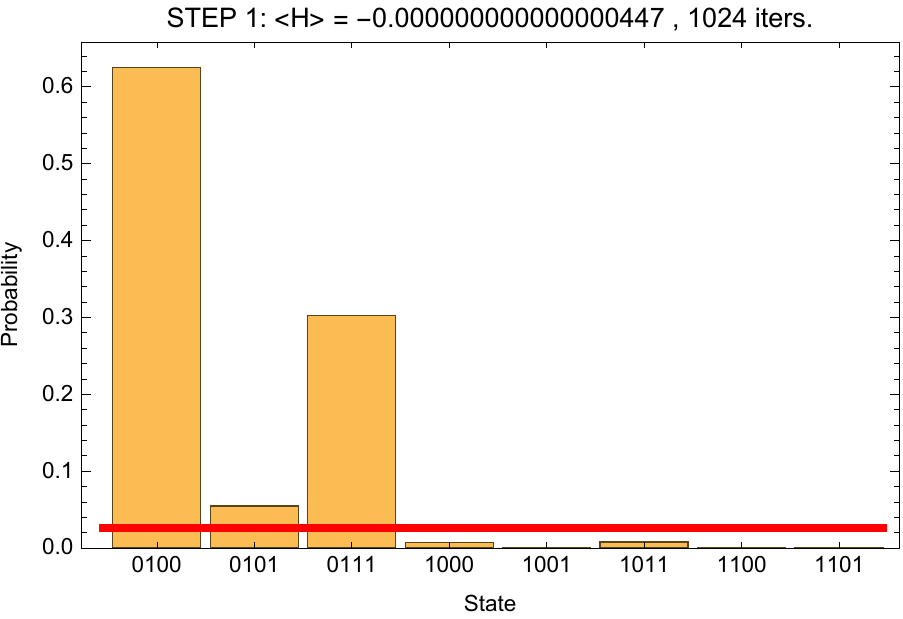} \ \
    \includegraphics[width=0.47\linewidth]{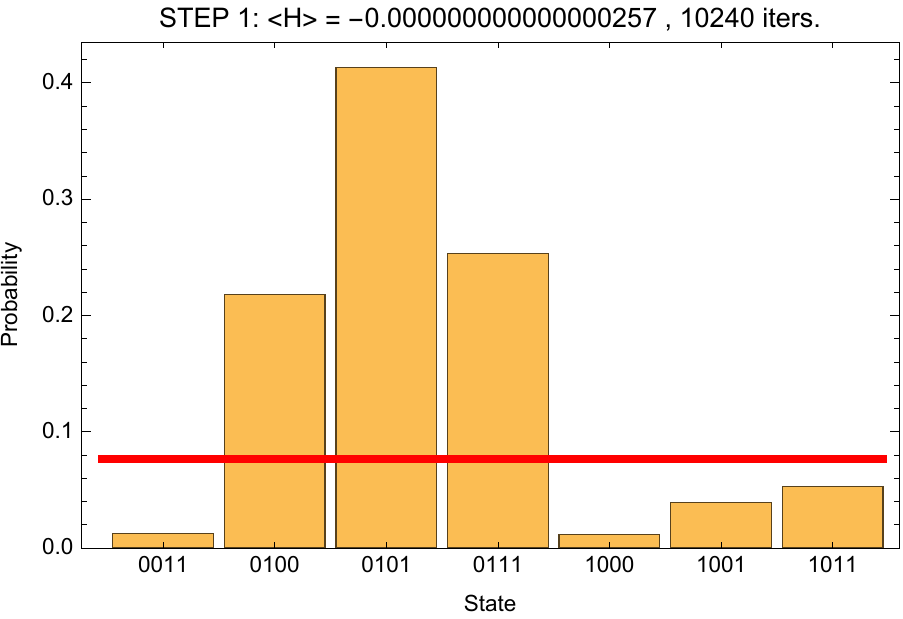}
    \caption{Approximation of the ground state found by VQE with \texttt{COBYLA} optimizer, corresponding to the two-eloop diagram in Fig. \ref{fig:TwoLoopsQC}. We present the probability of the different states contributing to the ground state after 1024 (left) and 10240 (right) iterations with 1024 shots. The red line is a threshold to limit the contamination by noncausal configurations.}
    \label{fig:ResultadoVQE0}
\end{figure*}
%++++++++++++++++++++++++++++++++++++++++++++++++++++++++++++++++++++++++++++++++++++

%%%%%%%%%%%%%%%%%%%%%%%%%%%%%%%%%%%%%%%%%%%%%%%%%%%%%%%%%%%%%%%%%%%%%%%%%%%%%%
%%%%%%%%%%%%%%%%%%%%%%%%%%%%%%%%%%%%%%%%%%%%%%%%%%%%%%%%%%%%%%%%%%%%%%%%%%%%%%
\subsection{Example: two-eloop case}
\label{ssec:2eloopHam}
Now, we concentrate again our attention to the explicit example of Fig.~\ref{fig:TwoLoopsQC}. As a first step, we proceed to translate \Eq{eq:HamiltonianoERestricted} to the basis of Pauli operators using the definition of the projectors given in \Eq{eq:Projector1}. The result is
\begin{align}
    H_{E/\{e_0\}} 
    &=  4 \, I_4\otimes I_3\otimes I_2\otimes I_1 
    + 2 \, I_4\otimes  I_3\otimes I_2\otimes Z_1 \nn\\
    &- I_4\otimes I_3\otimes Z_2\otimes I_1 
    - I_4\otimes I_3\otimes Z_2\otimes Z_1 \nn\\
    &+ I_4\otimes Z_3\otimes I_2\otimes I_1 
    + I_4\otimes Z_3\otimes I_2\otimes Z_1 \nn\\
    &+ 2 \, I_4\otimes Z_3\otimes Z_2\otimes I_1 
    + Z_4\otimes I_3\otimes I_2\otimes I_1 \nn\\
    &+ Z_4\otimes I_3\otimes I_2\otimes Z_1 
    + 2 \, Z_4\otimes I_3\otimes Z_2\otimes I_1 \nn\\
    &+ 3 \, Z_4\otimes Z_3\otimes I_2\otimes I_1 
    + Z_4\otimes Z_3\otimes I_2\otimes Z_1  ~. 
\label{eq:HamiltonianoERestrictedPAULI}
\end{align}
Then, we setup the VQE using the package \texttt{qiskit.algorithms}\footnote{\texttt{http://qiskit.org/}} together with the function \texttt{RealAmplitudes} to generate the Ans\"{a}tze. In this experiment, we use the optimizer \texttt{COBYLA}~\cite{Powell1994}, with \texttt{maxiter=$\{$1024,10240$\}$}, and the simulation is implemented in the \texttt{aer$\_$simulator} with a fixed number of shots (by default \texttt{shots=1024}). Notice that \emph{iterations} refers to the number of times that the classical optimization algorithm is called. This has to be distinguished from \emph{shots}, which is the number of times the quantum circuit is executed and measured. In any case, we observe that more iterations allow to achieve a better convergence of the optimization whilst more shots are associated with a more precise determination of the mean value of the Hamiltonian.

\begin{figure*}
    \centering
    \includegraphics[width=0.47\linewidth]{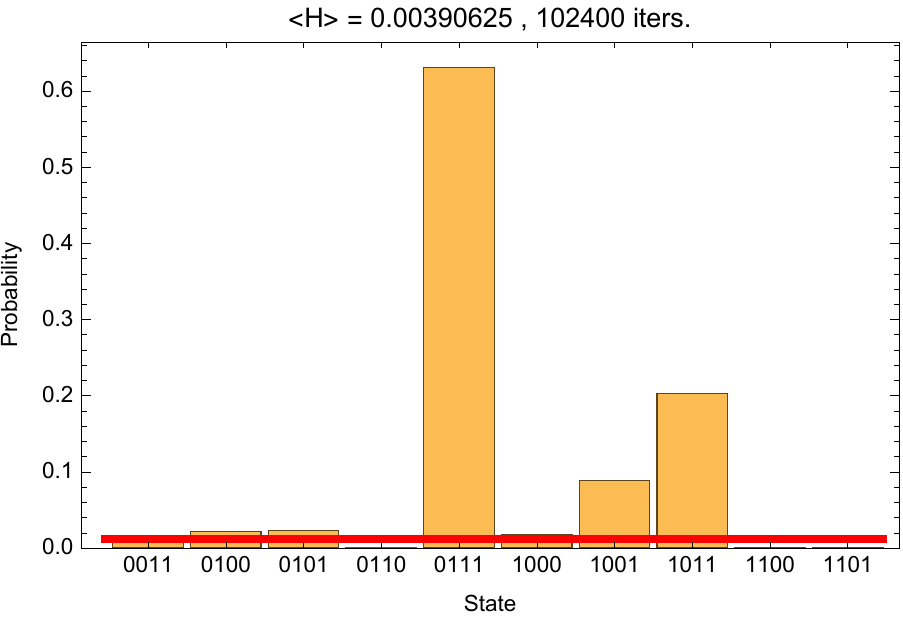} \ \
    \includegraphics[width=0.47\linewidth]{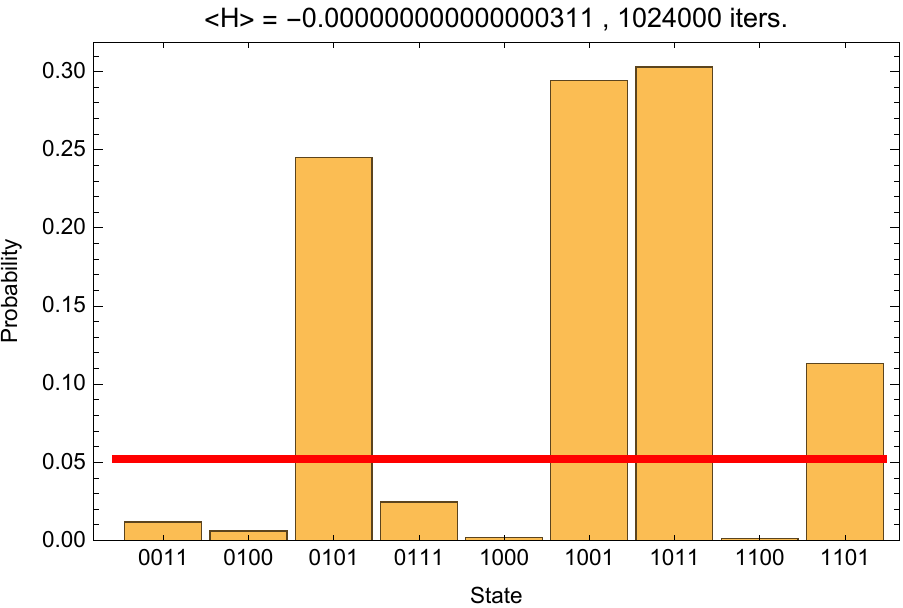}
    \caption{Approximation of the ground state found by VQE with \texttt{COBYLA} optimizer, corresponding to the two-eloop diagram in Fig. \ref{fig:TwoLoopsQC}. We present the probability of the different states contributing to the ground state after 102400 (left) and 1024000 (right) iterations, with 1024 shots. The red line is a threshold to limit the contamination by noncausal configurations.}
    \label{fig:ResultadoVQE1}
\end{figure*}

In Fig.~\ref{fig:ResultadoVQE0}, we show the probability of the different configurations contributing to the approximate ground state. The yellow bars are associated to acyclic (or causal) configurations, that the algorithm is aimed to identify. For the present topology, there are 9 acyclic configurations in the reduced space, namely:
\begin{align}
\big\{ &\ket{0011},\ \ket{0100},\ \ket{0101} ,\  \ket{0111},\ \nn\\[3pt]
&\ket{1000} ,\ \ket{1001},\  \ket{1011} ,\   \ket{1100} ,\  \ket{1101}  \big\} \, .
\end{align}
In this particular example, the simulation leads to the identification of 3 solutions, for both \texttt{maxiter}$=1024$ and \texttt{maxiter}$=10240$. The extremely good approximation to the exact energy of the ground state implies in practice that there was no contamination with noncausal configurations. However, this is not guaranteed, as we will see in subsequent examples along the article.
Furthermore, we notice that a single VQE \emph{run} (i.e., the set of quantum and classical steps discussed so far, which includes a certain number of shots and iterations) is generally not enough to collect all the solutions of the problem. Since there are several local minima, the minimization algorithm could be stuck in one of them and achieve convergence without exploring the full set of possibilities. The situation is even more complicated because there are multiple degenerated global minima, whose contribution to the ground state can be re-scaled arbitrarily without changing the energy. This implies that the random selection of the initial points might produce very different probability distributions from run to run: in other words, if we re-run the experiment, it is highly probable that the distributions in Fig. \ref{fig:ResultadoVQE0} change substantially, selecting a different subset of causal solutions. For these reasons, we define the \emph{success rate} as
\beq 
r_{\rm success} = 
\frac{\#(\rm{total \ detected})}
{\#(\rm{causal \ total})\times \big(1 + \#(\rm{incorrect \ states})\big)} \,  ,
\label{eq:SuccessRate}
\eeq
which also takes into account the possibility of misidentification of causal states. For the particular example presented in Fig.~\ref{fig:ResultadoVQE0}, we have $r_{\rm success} = 0.89$ and $r_{\rm success} = 0.78$, respectively. If we increase the number of iterations, we find the results in Fig. \ref{fig:ResultadoVQE1}: we have $r_{\rm success} = 0.56$ for \texttt{maxiter}$=102400$ and $r_{\rm success} = 1$ for \texttt{maxiter}$=1024000$ iterations. In the first case, one of the states, $\ket{0110}$, is noncausal, which leads to the drastic decrease of the success rate\footnote{Alternative definitions for the \emph{success rate} could be examined. Here we stick to this one in order to penalize the misidentification of causal states.}.

%++++++++++++++++++++++++++++++++++++++++++++++++++++++++++++++++++++++++++++++++++++
%++++++++++++++++++++++++++++++++++++++++++++++++++++++++++++++++++++++++++++++++++++

With the purpose of reducing the probability of collecting incorrect states, we can select only those states contributing to the approximated ground state with a probability above a certain threshold $\lambda$. Since the VQE is optimized to find the states with lower energy, it is reasonable to assume that there are much higher probabilities to include acyclic rather than noncausal configurations in the approximated ground state. Thus, if the solution found by VQE is
\beq
\ket{\psi_0} = \sum_j c_j \ket{\phi_j} \, ,
\eeq
then we propose the threshold $\lambda$ to be
\beq
\lambda = \max\left(\, \overline{(|c_j|^2)} -\frac{ \Delta (|c_j|^2)}{2}, \min(|c_j|^2) \, \right) \, , 
\label{eq:ThrDefinition1}
\eeq
i.e. the average minus a term proportional to the standard deviation of the probabilities $|c_j|^2$, or the minimum of such probabilities. This value of $\lambda$ corresponds to the red line in Figs. \ref{fig:ResultadoVQE0} and \ref{fig:ResultadoVQE1}. We would like to emphasize that this definition is empirical, and motivated by the exploration of the numerical results. Also, if we want to increase the certainty that only causal states are selected, we can take higher values of $\lambda$. This could eventually lead to a lower success rate (since more states are rejected). To balance this situation, we can run multiple times the VQE procedure and collect solutions in each step until we fulfill a certain convergence criterion, discussed in the next section.

To conclude this section, we would like to mention that different versions of the optimizers and Ans\"{a}tze were used. In particular, besides \texttt{COBYLA}, we performed the optimization with \texttt{SPSA}~\cite{spall1998overview} and \texttt{NFT}~\cite{Nakanishi_2020}, achieving similar results. However, \texttt{COBYLA} turned out to run faster on our computer\footnote{We are using a local version of \texttt{Qiskit} running on a Intel Core i5 desktop computer with 16 Gb of RAM memory.}. Regarding the Ansatz, we relied on \texttt{RealAmplitudes} and \texttt{EfficientSU2}, both circuits being part of \texttt{qiskit.circuit.library}. Again, the success rates for the topology A of Fig.~\ref{fig:TwoLoopsQC} were rather similar, although \texttt{RealAmplitudes} involves half of the parameters of \texttt{EfficientSU2}. However, the use of \texttt{RealAmplitudes} manifests into slightly higher values of the Hamiltonian mean value. Moreover, as we will see in the next section, it was necessary at times to switch to \texttt{EfficientSU2} to avoid missing solutions, while also avoiding misidentifications.

%%%%%%%%%%%%%%%%%%%%%%%%%%%%%%%%%%%%%%%%%%%%%%%%%%%%%%%%%%%%%%%%%%%%%%%%%%%%%%
%%%%%%%%%%%%%%%%%%%%%%%%%%%%%%%%%%%%%%%%%%%%%%%%%%%%%%%%%%%%%%%%%%%%%%%%%%%%%%
\subsection{Multiple-run VQE approach}
\label{ssec:IterativeVQE}
In order to find the complete set of solutions, one can set up a sequence of VQE runs where the Hamiltonian used contains additional terms which suppress the solutions found at each previous step of the sequence. The proposed \emph{theoretical} algorithm is the following:
\begin{enumerate}
        \item We start with the reduced Hamiltonian $H^{(0)} = H_{E/\{e_0\}}$ and apply the VQE algorithm using an Ansatz. We can choose as an initial point the uniform superposition of all the configurations, i.e.
        \beq 
        \ket{\psi^{(0)}} = \sqrt{2^{1-|E|}} \, \sum_i \ket{\phi_i} \, ,
        \eeq
        with $\{\ket{\phi_i}\}$ the canonical basis of the reduced edge space ${\cal H}_{E/\{e_0\}}$.
        \item The VQE minimization is executed till it finds a state with energy $E_1 < \tilde{\mu}$ or until the maximum number of iterations, \texttt{maxiter}, is reached\footnote{Here, we define $E_a=\bra{\psi_a} H \ket{\psi_a}$, i.e. we say that the energy of the state $\ket{\psi_a}$ is the expectation value of the Hamiltonian in that state.}. In case the energy condition is not fulfilled, we run the VQE again till it converges to $E_1 < \tilde{\mu}$ (which is guaranteed by the presence of, at least, 1 causal configuration). The approximation to the ground state is expressed as
        \beq 
        \ket{\psi^{(1)}} = \sum_j c_j^{(1)} \ket{\phi_j} \, .
        \eeq
        \item We pick up the subset ${\cal S}_1=\{\ket{\phi_j} \, | \,  |c_j^{(1)}|^2 > \lambda \}$, which corresponds to elements with a high probability of belonging to $\Ker H_{E/\{e_0\}}$. Then, we define a \emph{penalization term} to exclude these potential solutions from the next run. Explicitly, we consider
        \beq
        \Pi^{(1)} = \sum_{\ket{\phi_l} \in {\cal S}_1} b^{(1)}_l \ket{\phi_l}\bra{\phi_l} \, ,
        \eeq
        with $b_l^{(1)}$ positive coefficients to be set appropriately. Steps 1-3 define the single-run VQE, used in Sec.~\ref{ssec:2eloopHam}.
        \item We define the modified reduced Hamiltonian
        \beq 
        H^{(1)} = H^{(0)} + \Pi^{(1)} \, ,
        \label{eq:penalization_term}
        \eeq
        and re-run the VQE. If it does not find any state such that $E_2 < \tilde{\mu}$ within the given allowed number of iterations, then the algorithm stops. The set of solutions remains ${\cal S}={\cal S}_1$.
        \item If a state  
        \beq 
        \ket{\psi^{(2)}} = \sum_j c_j^{(2)} \ket{\phi_j} \, ,
        \eeq
        such that $E_2 < \tilde{\mu}$ is found, then we pick up the set ${\cal S}_2=\{\ket{\phi_j} \, | \,  |c_j^{(2)}|^2 > \lambda \}$, build the associated penalization term $\Pi^{(2)}$ and add ${\cal S}_2$ to the set of solutions.
        \item Repeat Step 4 and 5 till the algorithm stops when the mean energy is greater than $\tilde{\mu}$. 
\end{enumerate}
This approach is rather general and can be modified by adjusting several parameters: we can choose different energy thresholds $\tilde{\mu}$ for each step, different Ans\"{a}tze, probability thresholds, modify the number of shots and/or classical iterations for each VQE run, etc.. Although in the next section, we will discuss some of these observations, we defer a more detailed study for future works. Also, we would like to point out that this procedure does not guarantee collecting all the solutions. Due to the presence of so-called barren plateaus~\cite{mcclean2018barren}, the VQE could get stuck around local minima which do not belong to the true ground-state. This is a shared drawback of several minimization strategies (including VQE), in which the presence of large regions in the parameter space with rather small gradients prevents an efficient optimization~\cite{sack2022avoiding}. In the next section, we discuss improvements to overcome barren plateaus and increase the success rate, keeping a very low error rate.

Also, in the VQE literature \cite{fedorov2021vqe}, it is normally highlighted the importance of choosing a suitable Ansatz that properly represents at least the low-lying eigenstates of the Hamiltonian. In the context of quantum chemistry, this is usually possible since the Hamiltonians represent physical systems and analytical approximations or mathematical properties of the solutions are known. By choosing an appropriate Ansatz it would be possible to achieve a faster convergence and partially avoid the barren plateaus. However, in our case, the situation is not straightforward and identifying an optimal Ansatz is far from trivial. The choice of the Ansatz in general should have parametrized rotational gates and one or more layers of entanglement to be able to connect the qubits, if the entanglement entropy of the system is not zero. So, for finding convergence in this particular problem, it is enough to have a generic circuit structure. Our approach can be considered as a \emph{first exploration}, and further investigations in this direction will be performed in the future.

%++++++++++++++++++++++++++++++++++++++++++++++++++++++++++++++++++++++++++++++++++++
\begin{figure*}
    \centering
    \includegraphics[width=0.47\linewidth]{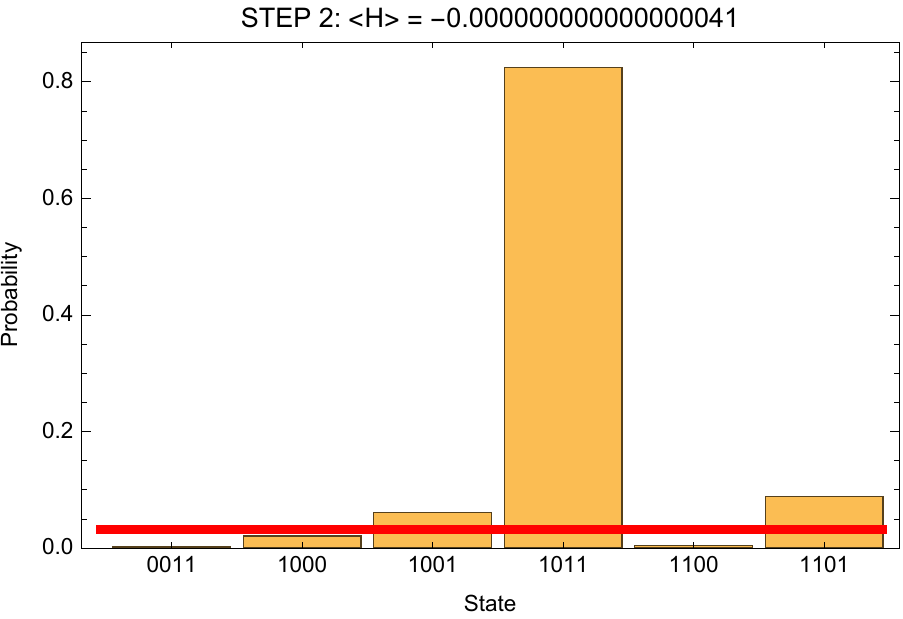} \ \
    \includegraphics[width=0.47\linewidth]{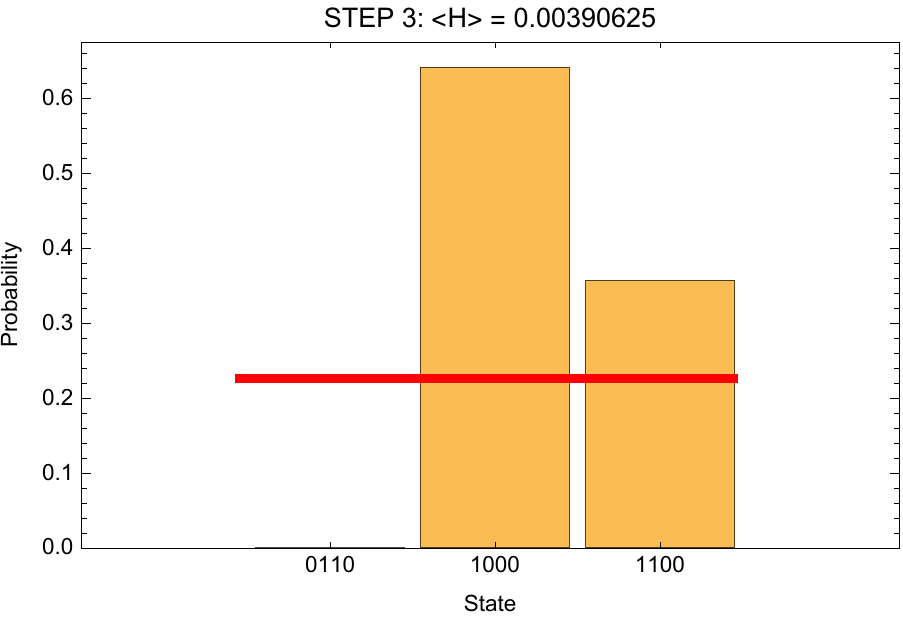} \ \
    \includegraphics[width=0.47\linewidth]{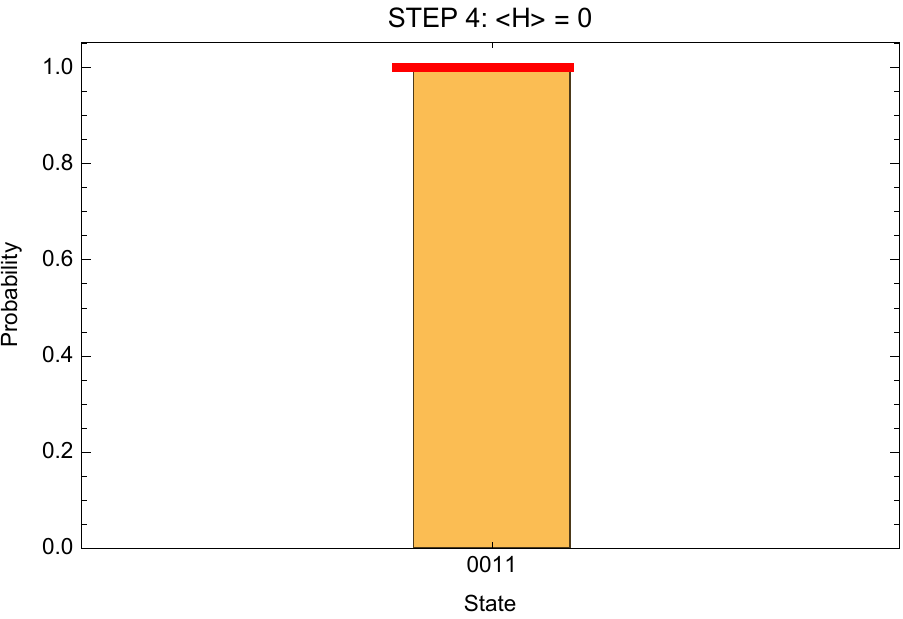} \ \
    \includegraphics[width=0.47\linewidth]{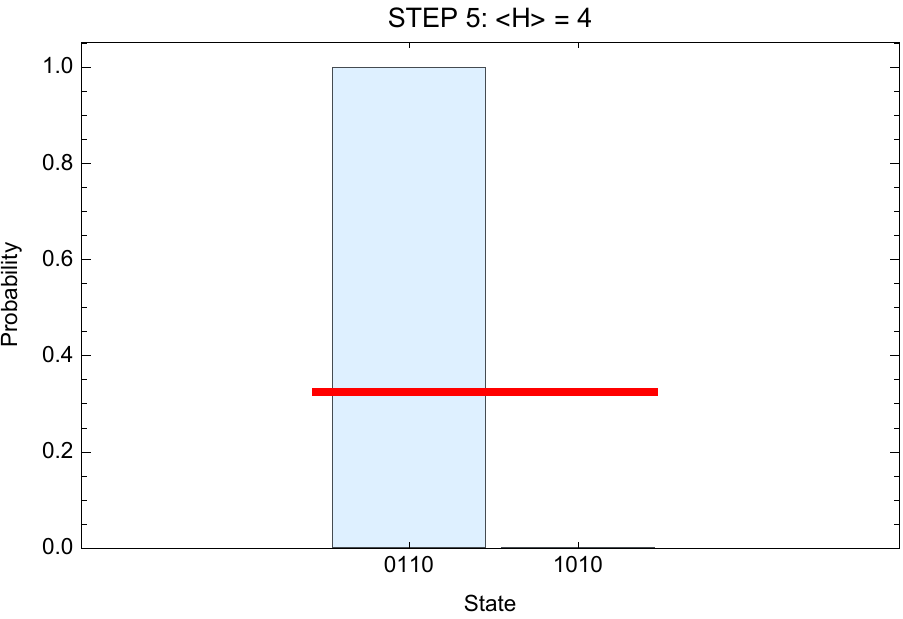}
    \caption{Iterated application of the VQE with \texttt{COBYLA} optimizer, corresponding to the two-eloop diagram in Fig. \ref{fig:TwoLoopsQC}. We present the probability of the different states contributing to the ground state after 1024 iterations. States in yellow (light-blue) represent causal (noncausal) solutions. The red line is a threshold to limit the contamination by noncausal configurations, described in the text. The algorithm converges after 4 runs.}
    \label{fig:ResultadoVQE1024}
\end{figure*}
%++++++++++++++++++++++++++++++++++++++++++++++++++++++++++++++++++++++++++++++++++++

%++++++++++++++++++++++++++++++++++++++++++++++++++++++++++++++++++++++++++++++++++++
\begin{figure*}
    \centering
    \includegraphics[width=0.47\linewidth]{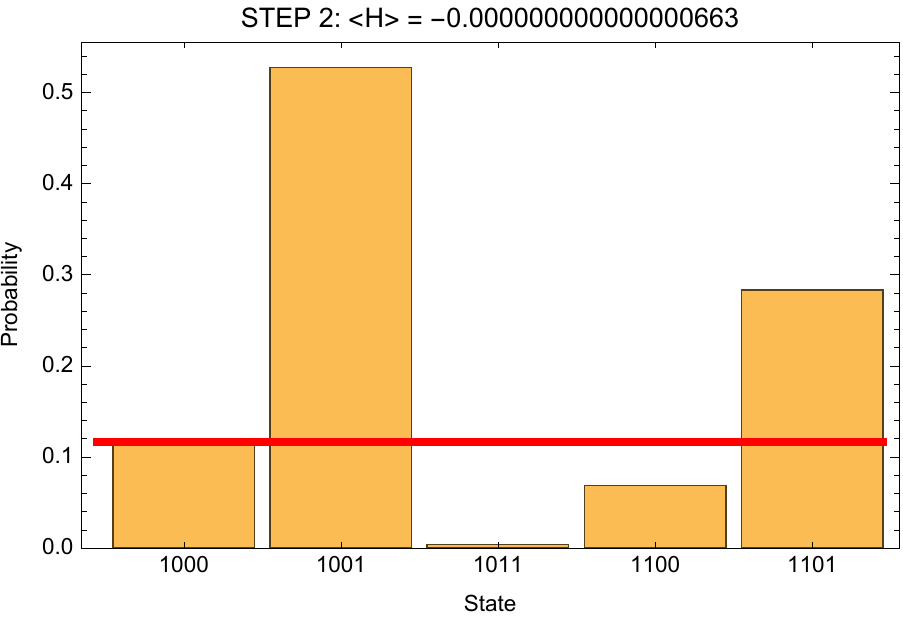} \ \
    \includegraphics[width=0.47\linewidth]{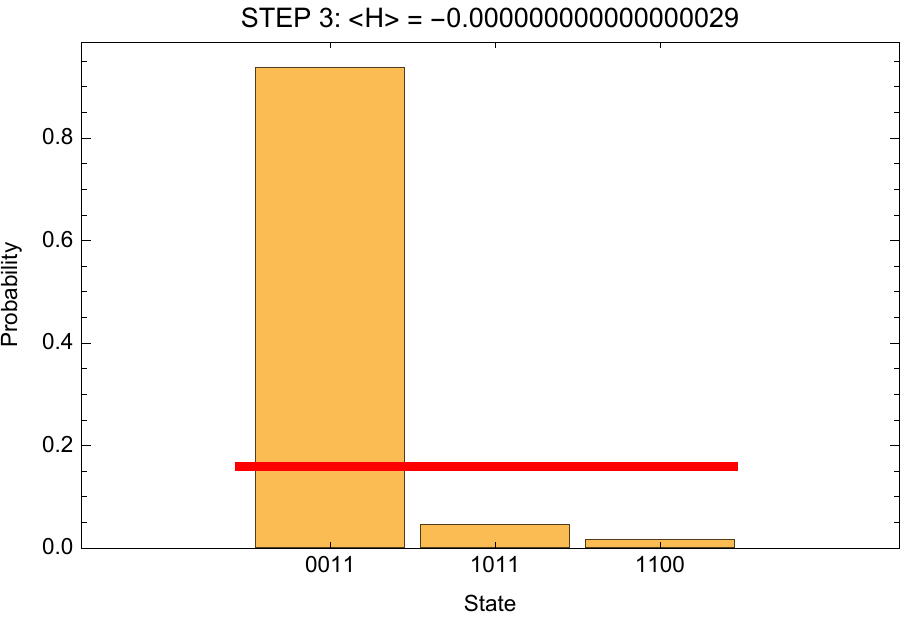} \ \ 
    \includegraphics[width=0.47\linewidth]{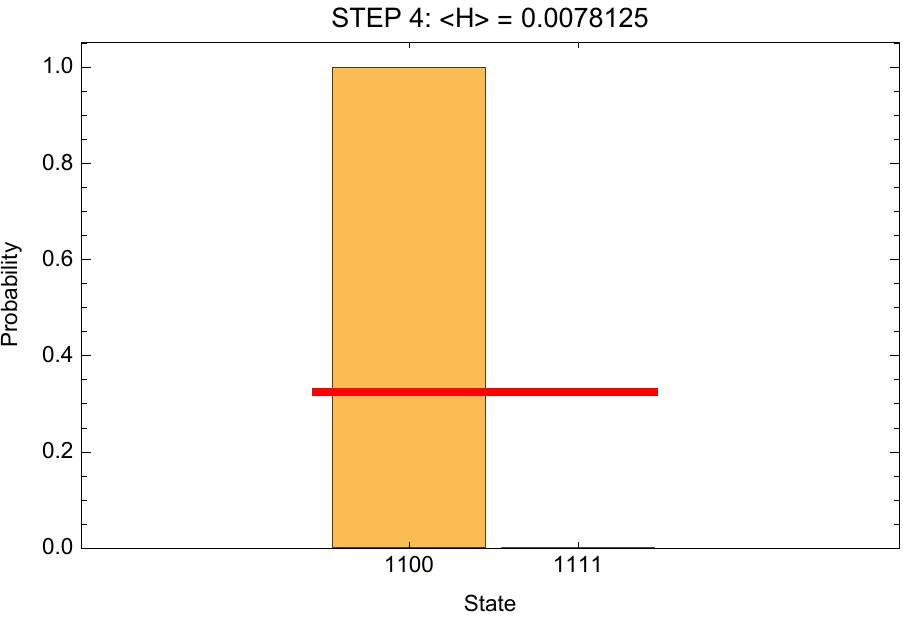} \ \
    \includegraphics[width=0.47\linewidth]{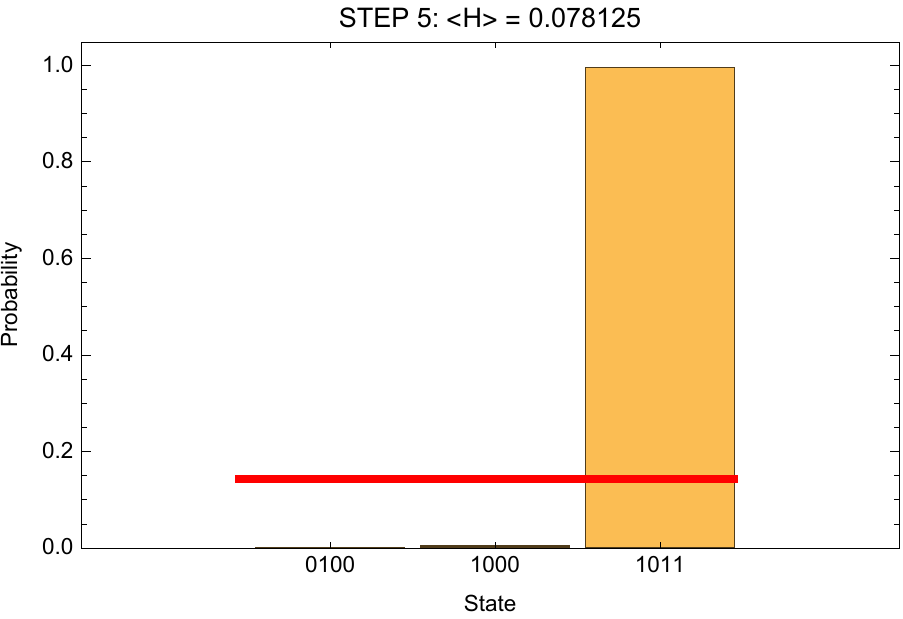} \ \
    \includegraphics[width=0.47\linewidth]{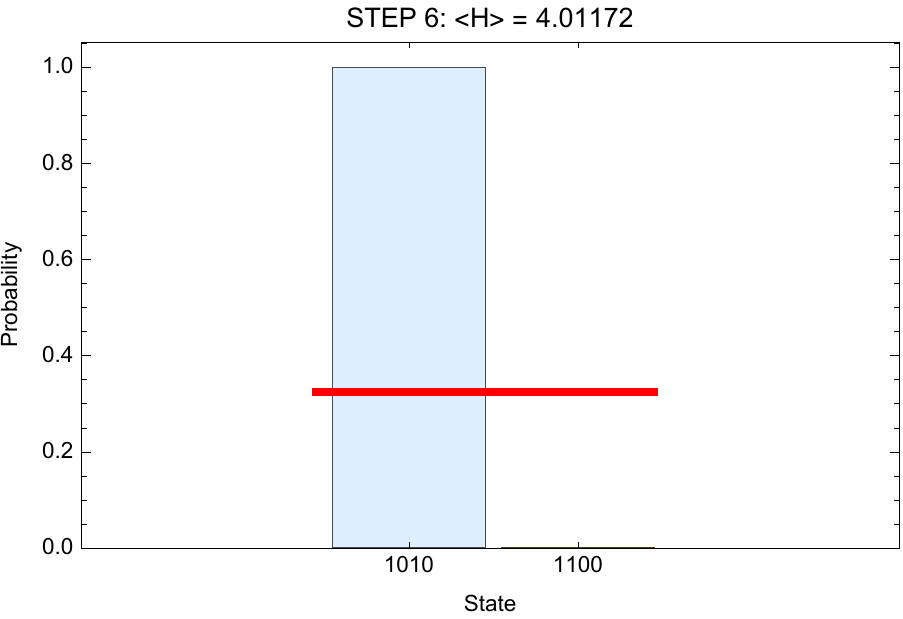}
    \caption{Iterated application of the VQE with \texttt{COBYLA} optimizer, corresponding to the two-eloop diagram in Fig. \ref{fig:TwoLoopsQC}. We present the probability of the different states contributing to the ground state after 10240 iterations. States in yellow (light-blue) represent causal (noncausal) solutions. The red line is a threshold to limit the contamination by noncausal configurations, described in the text. The algorithm converges after 5 runs.}
    \label{fig:ResultadoVQE10240}
\end{figure*}
%++++++++++++++++++++++++++++++++++++++++++++++++++++++++++++++++++++++++++++++++++++

Now, let us consider again the example of Fig. \ref{fig:TwoLoopsQC}, keeping the two test scenarios (i.e. the ones with 1024 and 10240 iterations, respectively). The starting point is the set of solutions collected from Fig. \ref{fig:ResultadoVQE0}. We set the detection threshold $\lambda$ according to Eq. (\ref{eq:ThrDefinition1}), and we iterate the VQE excluding the solutions found in the precedent step, as in Eq.~\eqref{eq:penalization_term}. In Figs. \ref{fig:ResultadoVQE1024} and \ref{fig:ResultadoVQE10240}, we present the probabilities of the different states contributing to the approximated ground state, for 1024 and 10240 iterations respectively. In each run, we collect the solutions above the threshold only if the energy is below $\tilde{\mu}=0.1$. In both cases, we manage to successfully identify the 9 causal states, after 5 and 6 runs, for \texttt{maxiter}$=1024$ and \texttt{maxiter}$=10240$ respectively. No misidentification took place because the selection threshold turned out to be rather rigorous. By lowering $\lambda$, the procedure converges faster, but there is an increased risk to pick up a noncausal state.

Before concluding this section, let us briefly discuss the complexity of this approach. Again, the multi-run VQE strategy involves a hybrid classical-quantum setup. On one side, there is a polynomial computational cost for calculating the Hamiltonian, which is ${\cal O}(|V|^4)$ based on Eq. (\ref{eq:loopHam_trn}), i.e., distinct $ |V| $ rows and $ |V| $ columns have to be multiplied, at a cost of $ |V| $ per multiplication, and this has to be repeated a total of $ |V|-1 $ times when $ n = |V| $.
Then, this Hamiltonian has to be evaluated in a quantum system, leading to a potential speed-up since all the expectation values (i.e., one for each term; see, e.g., \Eq{eq:HamiltonianoERestrictedPAULI}) are simultaneously computed on a superposition of many configurations \cite{Tilly:2021jem}. After that, a classical parameter optimization is performed with standardised algorithms (such as \texttt{COBYLA} or \texttt{NFT}) and the procedure is iterated.

%%%%%%%%%%%%%%%%%%%%%%%%%%%%%%%%%%%%%%%%%%%%%%%%%%%%%%%%%%%%%%%%%%%%%%%%%%%%%%
%%%%%%%%%%%%%%%%%%%%%%%%%%%%%%%%%%%%%%%%%%%%%%%%%%%%%%%%%%%%%%%%%%%%%%%%%%%%%%
\section{Comparing Hamiltonian optimization with Grover's algorithm}
\label{sec:ConnectionGH}
\begin{figure}[h!]
    \centering
    \includegraphics[width=\linewidth]{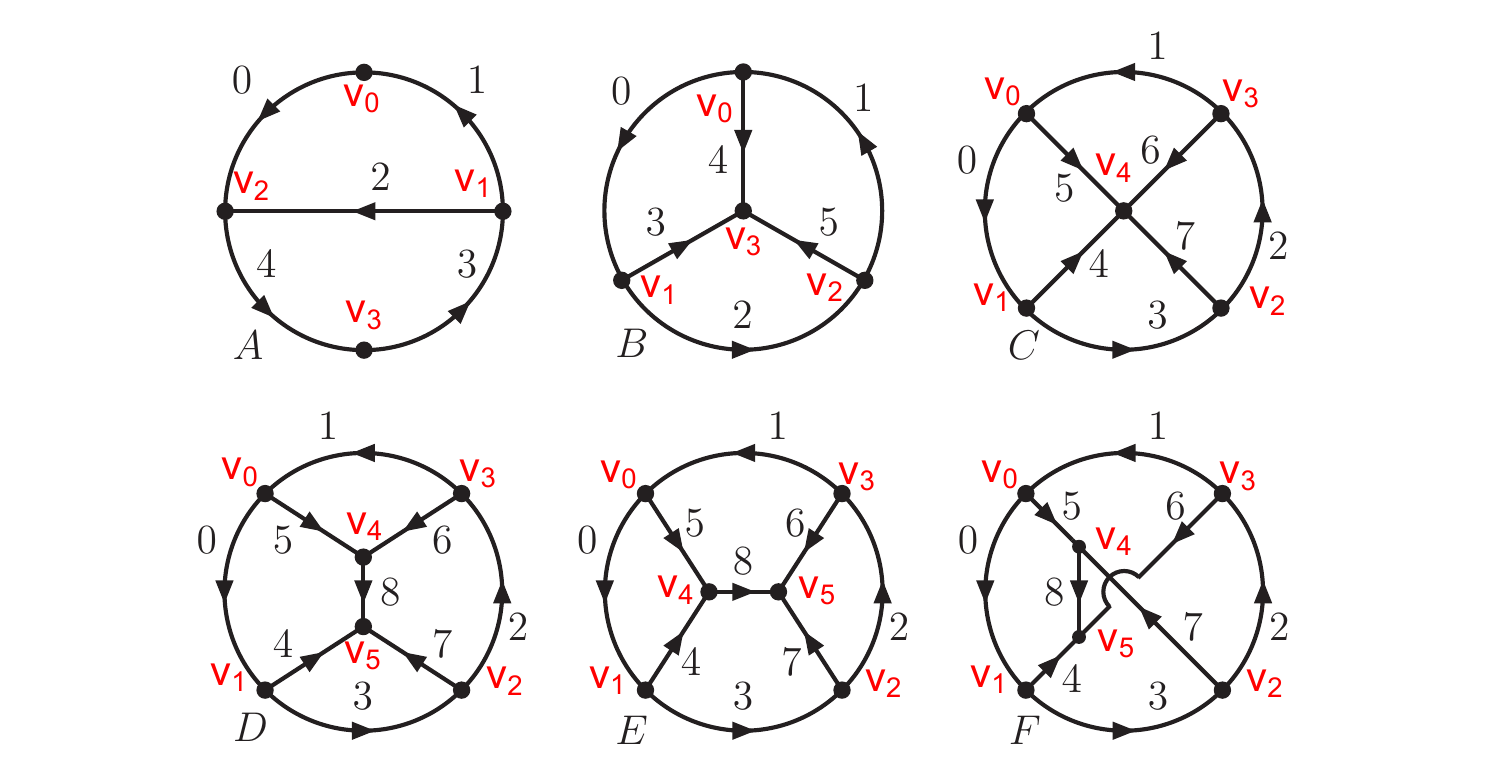}
    \caption{Representative topologies used for the performance comparison of Grover's and VQE implementations. We considered the following configurations: (A) two eloops with four vertices; (B) three eloops with four vertices; and (C) four eloops with five vertices. Also, we include the complete set of four-eloop six-vertex topologies: (D) $t$-channel; (E) $s$-channel; and (F) $u$-channel.
    \label{fig:digraph6}}
\end{figure}

After a careful study of the Hamiltonian minimization through VQE, we carry out in this Section a performance comparison with Grover's based algorithm~\cite{Ramirez-Uribe:2021ubp}. We rely on the set of six representative topologies shown in Fig. \ref{fig:digraph6}. These topologies allow us to understand how the quantum circuits' complexity scales with the number of vertices and edges. In Appendix \ref{app:Hamiltonians}, we show the explicit form of the Hamiltonians in $\mathcal{H}_E$ for the aforementioned topologies, so that a classical verification of the procedure explained in Sec.~\ref{sec:Hamiltonian} can be implemented.

%++++++++++++++++++++++++++++++++++++++++++++++++++++++++++++++++++++++++++++++++++++
%++++++++++++++++++++++++++++++++++++++++++++++++++++++++++++++++++++++++++++++++++++

For each topology, we study the set of causal/acyclic configurations comparing with the output of a classical algorithm\footnote{We rely on a \emph{naive} implementation that checks whether or not the adjacency matrix is nilpotent for a given graph.}. We used the success rate defined in \Eq{eq:SuccessRate} to quantify the accuracy of the identification of causal states. Then, we compare:
\begin{enumerate}
    \item the number of qubits required to run the algorithm;
    \item the quantum depth of the \emph{transpiled} circuits;
    \item the \emph{total number of executions} needed to get the full solution.
\end{enumerate}
The first point is important since most of the current quantum devices have a (relatively) small number of qubits. Of course, we expect that this limitation will be overcome in the (near) future, but still it is a tangible drawback. In the same direction, physical bottlenecks to ensure the stability of the qubits prevents to implement very deep circuits. So, from the point of view of the hardware requirements, avoiding a fast growth of these two aspects is crucial to ensure the feasibility of the quantum approach. Finally, the third point implies more execution time, since re-running the circuit implies preparing the state and repeating the measurements. However, this is not a hard limiting factor as the previous two properties, although it penalizes the performance.

A further comment is needed to explain the definition of \emph{quantum depth} and \emph{executions} in this comparison. Regarding the \emph{quantum depth}, it is related to the number of operations or quantum gates that are involved within the circuit. Each specific device has a set of fundamental gates, and any other gate/operator is defined by a combination of the fundamental set. As a consequence, the real depth of the circuit is hardware- or simulator- dependent, so we need to fix a convention to compare Grover's and VQE approaches. Since the calculations are implemented in \texttt{Qiskit}, we use the function \texttt{qiskit.compiler.transpile} and \texttt{aer$\_$simulator} as target device. Then, we estimate the quantum depth through the \texttt{.depth()} method.  For the case of Grover's algorithm, the depth depends not only on the complexity of the topology but also on the number of \emph{iterations}: for the topologies studied in this article, we can achieve a perfect reconstruction with only one iteration \cite{Ramirez-Uribe:2021ubp}. Note that in this case, the complexity of the oracle and the diffuser are included. The quantum depth for VQE is related to the complexity of the Ansatz and the Hamiltonian. In fact, for the VQE, we estimate the depth by applying the \texttt{.depth()} method to the circuit composed by one term of the Hamiltonian applied to the Ansatz. So, in general, VQE involves \emph{shorter} circuits, as discussed in e.g. \cite{Tilly:2021jem}.

About the concept of \emph{executions}, it is directly related to the number of times that we need to measure the output. In the case of Grover's algorithm, this quantity exactly agrees with the number of shots \cite{Grover:1997ch,Ramirez-Uribe:2021ubp}. However, VQE requires a deeper analysis. We need to remember that the VQE pipeline mixes a classical and a quantum algorithm. The quantum circuits depend on parameters, that are adjusted after performing the measurement in order to minimise a certain cost function~\cite{Tilly:2021jem}. For each choice of parameters, we define a quantum circuit and execute $s$ shots (controlled by the parameter \texttt{shots} within the backend). Then, we allow the algorithm to modify the parameters $r$ times (until the cost function evaluates under a certain threshold), which is what we call an \emph{iteration} (controlled by \texttt{maxiterations} within the classical optimizer routine). So, each \emph{run} of the VQE requires $s\times r$ measurements. On top of that, the multiple-run VQE algorithm explained in Sec.~\ref{sec:NumericalHamiltonian} must be repeated $k$ times, adding in each repetition a certain number of penalization terms to the Hamiltonian. As a consequence, we will need to make $s \times r \times k$ \emph{executions} to complete the identification of the winning states (or causal configurations). For these reasons, in general, VQE involves a \emph{larger} number of executions compared to Grover's approach.  

%++++++++++++++++++++++++++++++++++++++++++++++++++++++++++++++++++++++++++++++++++++
\begin{table*}
\begin{center}
\begin{tabular}{cccccc} \hline \hline 
Topology & Causal/Total & Success rate & $\#$ Qubits & Quantum depth & $\#$ Executions \\ \hline
A & 18/32   & 100 $\%$  & 14    & 35    & 100   \\
B & 24/64   & 100 $\%$  & 19    & 43    & 100   \\
C & 78/256  & 100 $\%$  & 25    & 55    & 400   \\
D & 204/512 & 100 $\%$  & 28    & 62    & 1300  \\
E & 204/512 & 100 $\%$  & 28    & 62    & 1300  \\
F & 230/512 & 100 $\%$  & 33    & 28     & 1600\\ 
\hline \hline 
\end{tabular}
\end{center}
\caption{Performance of the Grover-based algorithm applied to the representative multiloop topologies depicted in Fig.~\ref{fig:digraph6} \label{tb:RESOURCESGROVER}} 
\end{table*}
%++++++++++++++++++++++++++++++++++++++++++++++++++++++++++++++++++++++++++++++++++++

%++++++++++++++++++++++++++++++++++++++++++++++++++++++++++++++++++++++++++++++++++++
\begin{table*}
\begin{center}
    \begin{tabular}{cccccc} \hline \hline 
Topology & Causal/Total & Success rate & $\#$ Qubits & Quantum depth & $\#$ Executions \\ \hline
A & 18/32 & 60.7 - 79.1 $\%$ & 4 & 7 & 1000 $\times$ 100\\
B & 24/64 & 23.9 - 56.8  $\%$ & 5 & 9 & 1000 $\times$ 1000\\
C & 78/256 & 7.9 - 37.6 $\%$ & 7 & 13 & 1000 $\times$ 1000\\
D & 204/512 & 3.5 - 35.4 $\%$ & 8 & 15 & 10000 $\times$ 1000\\
E & 204/512 & 4.0 - 30.5 $\%$ & 8 & 15 & 10000 $\times$ 1000\\
F & 230/512 & 4.6 - 34.6 $\%$ & 8 & 15 & 10000 $\times$ 1000\\
\hline \hline 
\end{tabular}
\end{center}
\caption{Performance of the Hamiltonian minimization through VQE applied to the representative multiloop topologies depicted in Fig.~\ref{fig:digraph6}, using a single run. We indicate the number of executions as \emph{shots} times \emph{iterations}. The success rate $r_{\rm success}$ range contemplates the worst case scenario (no threshold, all the states are selected) and the best one (optimal threshold for distinguishing causal from noncausal). Thus, these results illustrate the need for defining a threshold to avoid misidentifications. \label{tb:RESOURCESVQE0}} 
\end{table*}
%++++++++++++++++++++++++++++++++++++++++++++++++++++++++++++++++++++++++++++++++++++

%++++++++++++++++++++++++++++++++++++++++++++++++++++++++++++++++++++++++++++++++++++
\begin{table*}
\begin{center}
    \begin{tabular}{cccccc} \hline \hline 
Topology & Causal/Total & Success rate & $\#$ Qubits & Quantum depth & $\#$ Runs \\ \hline
A & 18/32 & 88 $\%$ & 4 & 7 & 3\\
B & 24/64 & 50 $\%$ & 5 & 9 & 4\\
C & 78/256 & 51 $\%$ & 7 & 13 & 7\\
D & 204/512 & 34.3 $\%$ & 8 & 15 & 7\\
E & 204/512 & 43.1 $\%$ & 8 & 15 & 11\\
F & 230/512 & 44.3 $\%$ & 8 & 15 & 9\\
\hline \hline 
\end{tabular}
\end{center}
\caption{Performance of the Hamiltonian minimization through VQE applied to the representative multiloop topologies depicted in Fig.~\ref{fig:digraph6}, using the iterative approach explained in Sec.~\ref{sec:NumericalHamiltonian}. We used 1000 iterations and 1000 shots in each VQE run, with the \emph{Setup 2}. Even if the success rate was below 100 $\%$, no misidentification took place. \label{tb:RESOURCESVQEv2}} 
\end{table*}
%++++++++++++++++++++++++++++++++++++++++++++++++++++++++++++++++++++++++++++++++++++

%++++++++++++++++++++++++++++++++++++++++++++++++++++++++++++++++++++++++++++++++++++
\begin{table*}
\begin{center}
    \begin{tabular}{cccccc} \hline \hline 
Topology & Causal/Total & Success rate & $\#$ Qubits & Quantum depth & $\#$ Runs \\ \hline
A & 18/32 & 100 $\%$ & 4 & 9 & 3\\
B & 24/64 & 100 $\%$ & 5 & 11 & 9\\
C & 78/256 & 97.4 $\%$ & 7 & 15 & 18\\
D & 204/512 & 93.1 $\%$ & 8 & 17 & 41\\
E & 204/512 & 95.1 $\%$ & 8 & 17 & 40\\
F & 230/512 & 87.0 $\%$ & 8 & 17 & 37\\
\hline \hline 
\end{tabular}
\end{center}
\caption{Performance of the Hamiltonian minimization through VQE applied to the representative multiloop topologies depicted in Fig.~\ref{fig:digraph6}, using the iterative approach explained in Sec.~\ref{sec:NumericalHamiltonian}. We used 1000 iterations and 1000 shots in each VQE run, with the \emph{Setup 3}. A major improvement compared to \emph{Setup 2} was reached, keeping a null error rate. \label{tb:RESOURCESVQEv3}} 
\end{table*}
%++++++++++++++++++++++++++++++++++++++++++++++++++++++++++++++++++++++++++++++++++++

In Tables~\ref{tb:RESOURCESGROVER} and \ref{tb:RESOURCESVQE0}, we present the number of qubits, executions and the quantum depth of the circuits involved in the implementation of the Grover's algorithm and the single-run VQE approach, respectively, for all the topologies shown in Fig.~\ref{fig:digraph6}. In the case of Grover, the success rate is 100 $\%$ because we are using the optimal number of executions explored in Ref.~\cite{Ramirez-Uribe:2021ubp}, which guarantees in practice a perfect casual/noncausal discrimination, in spite of requiring more qubits and larger coherence. For the single-run VQE, we estimate the success rate by using \Eq{eq:SuccessRate}. We rely on \texttt{COBYLA} optimizer and \texttt{RealAmplitudes} Ansatz which represents less quantum resources with respect to Grover's implementation but involves more shots, classical iterations and the efficiency is noticeably smaller. This is because, in the experiments that we executed to obtain the inferior values in the success rate ranges of Tab. \ref{tb:RESOURCESVQE0}, we are collecting all the states in the approximate ground-state as solutions of our problem. However, if the energy is not 0 (or compatible with 0 within machine error), this ground-state might be contaminated with noncausal configurations; this will reduce the success rate since we are penalizing the misidentification with Eq. (\ref{eq:SuccessRate}). This illustrates the need for defining a proper threshold to retain only the \emph{true} minima from the approximated ground-state.

Then, we tested the performance of the multiple-run VQE algorithm. We considered different scenarios:
\begin{enumerate}
\item \emph{Setup 1}: \texttt{COBYLA} optimizer with 1000 iterations, 1000 shots, \texttt{RealAmplitudes} with a random parameter initialization as Ansatz and the threshold $\lambda$ given by Eq. (\ref{eq:ThrDefinition1}).
\item \emph{Setup 2}: \texttt{NFT} optimizer with 1000 iterations, 1000 shots, \texttt{RealAmplitudes} with a random parameter initialization as Ansatz. The selection threshold for the $i$-th run is given by
\beq
\lambda_i = \max\left(\, \overline{(|c^{(i)}_j|^2)} -\frac{ \Delta (|c^{(i)}_j|^2)}{2}, \frac{1}{\# {\cal S}_i} \, \right) \, , 
\label{eq:ThrDefinition1BIS}
\eeq
if $E_i > 10 ^ {-8}$, and $\lambda_i = 0$ otherwise (i.e. if the mean energy of the approximated ground-state is compatible with 0, then all the states are solutions).
\item \emph{Setup 3}: \texttt{NFT} optimizer with 1000 iterations, 1000 shots, \texttt{EfficientSU2} using the approximated solution found in the previous run as initial point of its subsequent run. The selection threshold is defined as in \emph{Setup 2}, and we repeat 3 times the execution of the VQE if $E_i > \tilde{\mu}$ in the $i$-th run.
\end{enumerate}
\emph{Setup 1} corresponds to the configuration tested in Sec. \ref{ssec:IterativeVQE}. We do not present a table with explicit results for complex topologies (such as D, E of F) because the success rates are rather low, even below ${\cal O}(10 \, \%)$. This is due to several withdraws of this naive implementation, namely some limitations of \texttt{COBYLA} optimizer and the initial Ansatz. For this reason, we tried other configurations, tweaking the parameters of the multi-run VQE to achieve higher success rates. In Tabs. \ref{tb:RESOURCESVQEv2} and \ref{tb:RESOURCESVQEv3}, we present the results for \emph{Setup 2} and \emph{Setup 3} respectively. We observe that \texttt{NFT} identifies approximations to the real ground-state with lower energy but with less configurations in superposition w.r.t. \texttt{COBYLA}. In fact, with \texttt{COBYLA} the identification rate could reach up to ${\cal O}(50 \%)$, but there are several incorrect states, which leads to very low success rates, see Eq.~\eqref{eq:SuccessRate}. In contrast, the combination of the \texttt{NFT} optimization and the improved selection threshold from Eq. (\ref{eq:ThrDefinition1BIS}) has an extremely low misidentification rate: in all the topologies tested, the error rate was absent, but each run collects a smaller number of solutions. The difference in the performance is due to the nature of the optimization algorithm. \texttt{NFT} is an optimization method specifically designed for quantum-classical hybrid algorithms based on parameterized quantum circuits, whilst \texttt{COBYLA} is a more general approach envisaged for problems where the derivative of the objective function is unknown. Since it assumes certain properties of the cost function, \texttt{NFT} allows to reach configurations with lower energy by performing fewer measurements. Even if we were unable to test the performance of these two approaches in real quantum devices, we expect \texttt{NFT} to perform better for the multi-run VQE since it is also claimed to be more effective in the presence of noise (as well as \texttt{SPSA}). Still, in general, it is not clear from first principles which is the best choice of optimizer and, most of the times, this is found by empirical exploration.

Still, in presence of barren plateaus, the implementation in \emph{Setup 2} stops before identifying all the possible solutions, since the algorithm gets stuck around local minima. To overcome this limitation, in \emph{Setup 3}, we consider a more flexible Ansatz and select an optimized value for the initial point. Also, we implemented a routine to \emph{randomly kick the parameters} when the energy is higher than $\tilde{\mu} = 1$ (which means a non-global minimum was found). In this way, we avoid a premature interruption of the multi-run VQE, which keeps collecting solutions, with null error rate.

%%%%%%%%%%%%%%%%%%%%%%%%%%%%%%%%%%%%%%%%%%%%%%%%%%%%%%%%%%%%%%%%%%%%%%%%%%%%%%
%%%%%%%%%%%%%%%%%%%%%%%%%%%%%%%%%%%%%%%%%%%%%%%%%%%%%%%%%%%%%%%%%%%%%%%%%%%%%%
\section{Conclusions and outlook}
\label{sec:Conclusions}
The computation of accurate theoretical predictions for current and future high-energy particle colliders requires efficient ways to deal with scattering amplitudes at higher orders in the perturbative expansion. In the present article, we put forward a quantum strategy to re-cast the selection of causal configurations of multiloop Feynman integrands into a minimization problem. For this purpose, we exploit the geometrical approach to the causal Loop-Tree Duality (LTD) representation of Feynman integrals \cite{Sborlini:2021owe} and relate causality with the identification of directed acyclic graphs \cite{Ramirez-Uribe:2021ubp}.

We rely on the concept of adjacency matrix in graph theory to define the so-called \emph{loop Hamiltonian}, whose different energy levels correspond to graph configurations with different number of cycles. By construction, the ground-state is directly associated to the subspace of states related to directed acyclic graphs. In this way, given a multiloop Feynman diagram, we fix a reference orientation of the internal propagators and then define the associated loop Hamiltonian. By construction, this loop Hamiltonian has mean value 0 when evaluated on acyclic states. Therefore by minimizing the energy, we automatically detect the subset of causal configuration.

After explaining the construction of the loop Hamiltonian, we proceed to solve with quantum minimization algorithms. We use the Variational Quantum Eigensolver (VQE) within the \texttt{Qiskit} framework to test the different configurations. In particular, we develop a multi-run VQE strategy that allows to achieve higher detection rates. The VQE approach is compared against the Grover's detection procedure implemented in Ref.~\cite{Ramirez-Uribe:2021ubp}. Our results indicate that even if VQE requires less quantum resources than Grover's algorithm (i.e. it involves fewer qubits and shorter circuits), still a larger number of executions must be performed in order to achieve high success rates. However, since VQE is a hybrid algorithm that employs classical optimization, it allows to use classical hardware and save quantum resources. Given the current limitations in quantum hardware, VQE approach allows to study complex multiloop topologies that would require a prohibitive number of stable qubits within Grover's implementation.

In conclusion, along this article, we have presented a proof-of-principle for the detection of causal configurations using a loop Hamiltonian implemented in VQE. We have demonstrated the feasibility of our approach, paving the road to achieve an efficient causal reconstruction of multiloop Feynman amplitudes through quantum algorithms using less hardware resources. We have presented a multiple-run VQE algorithm with much better success rates than a naive single-run VQE. We stress that we avoid the misidentification of solutions, reaching null error rates in the simulators. Still, the success rates achieved are not as satisfactory as the ones achieved with a Grover-based algorithm and further investigation is required to develop a better strategy through quantum minimization. In any case, we successfully tested, for the first time, a VQE-based strategy to solve a problem with highly-degenerated ground-states, which constitutes a hard stress-test for this kind of minimization algorithms. A promising road to enhance the success rate and reduce the number of VQE runs is Conditional-Value-at-Risk (CVaR)~\cite{barkoutsos2020improving}, which has already shown improved performance on certain practical problems, such as the allocation of flight gates~\cite{stollenwerk2019flight,Chai:2023ixt}. Also, different strategies must be explored regarding optimal state-preparation, since several results point towards a significant impact of the state-preparation in the performance of the minimization \cite{Bravyi_2020,Sun_2023}. In particular, it would be highly interesting to test our algorithms in quantum annealers, which could lead to an improved performance and much higher success rates with a lower consumption of quantum resources. We defer these improvements and studies for future investigations.

%%%%%%%%%%%%%%%%%%%%%%%%%%%%%%%%%%%%%%%%%%%%%%%%%%%%%%%%%%%%%%%%%%%%%%%%%%%%%%%%%
%%%%%%%%%%%%%%%%%%%%%%%%%%%%%%%%%%%%%%%%%%%%%%%%%%%%%%%%%%%%%%%%%%%%%%%%%%%%%%%%%
\section*{Acknowledgements}
We would like to thank Y. Chai, R. J. Hern\'andez-Pinto, S. K\"{u}hn, D. F. Renter\'ia-Estrada, S. Ochoa-Oreg\'on and P. Zurita for fruitful discussions about VQE and quantum algorithms for minimization problems. This work was supported by the Spanish Government (Agencia Estatal de Investigaci\'on MCIN/AEI/10.13039/501100011033) Grants No. PID2020-114473GB-I00 and PID2019-105439GB-C22, and Generalitat Valenciana Grant No. PROMETEO/2021/071 and ASFAE/2022/009. 
AC is supported in part by the Helmholtz Association - “Innopool Project Variational Quantum Computer Simulations (VQCS)”.
G.S. is partially supported by Programas Propios II (Universidad de Salamanca), EU Horizon 2020 research and innovation program STRONG-2020 project under grant agreement No. 824093 and H2020-MSCA-COFUND USAL4EXCELLENCE-PROOPI-391 project under grant agreement No 101034371. L.V.S. has received funding from the EU Horizon 2020 research and innovation programme under the Marie Sklodowska-Curie grant agreement No 101031558.
S.R.-U. acknowledges support from CONACyT and Universidad  Aut\'onoma  de  Sinaloa, and A.E.R.-O. from the Spanish Government (PRE2018-085925).

\appendix

%%%%%%%%%%%%%%%%%%%%%%%%%%%%%%%%%%%%%%%%%%%%%%%%%%%%%%%%%%%%%%%%%%%%%%%%%%%%%%
%%%%%%%%%%%%%%%%%%%%%%%%%%%%%%%%%%%%%%%%%%%%%%%%%%%%%%%%%%%%%%%%%%%%%%%%%%%%%%
\section{Illustrative application of geometrical causal rules}
\label{app:CAUSALEJEMPLO}
In Sec. \ref{sec:Causalflow}, we briefly mentioned how to determine the causal entangled thresholds by following geometrical selection rules. Let us consider the diagram shown in Fig.~\ref{fig:TwoLoopsQC} as an illustrative example of the bootstrapping of the causal representation. We will closely follow Refs.~\cite{Sborlini:2021nqu,Sborlini:2021owe}, together with the concepts exposed in Refs.~\cite{Verdugo:2020kzh,Aguilera-Verdugo:2020kzc,Aguilera-Verdugo:2021nrn,TorresBobadilla:2021ivx}.

The first step consists in the identification of the causal thresholds or causal propagators, which corresponds to connected binary partitions of the set of vertices. The diagram Fig.~\ref{fig:TwoLoopsQC} has $V=4$ vertices connected by $E=5$ edges, which produces a set of 6 causal propagators
\beqn
\lambda_{i+1} &=& \{v_i\} \quad {\rm for} \, i = \{ 0,\ldots,3\} \, ,
\\ \lambda_{5} &=& \{v_0,v_1\} \, ,
\\ \lambda_{6} &=& \{v_0,v_2\} \, .
\eeqn
Each $\lambda_p$ is graphically indicated by a line crossing the edges that enclose the selected vertices; we will call them \emph{causal cuts}, since they \emph{cut} the diagram into two connected ones. For instance, $\lambda_1$ corresponds to a line crossing the edges $\{e_0,e_1\}$ and enclosing the vertex $v_0$. Furthermore, as explained in Refs. \cite{Verdugo:2020kzh,Sborlini:2021owe}, we define $\lambda_p^+$ ($\lambda_p^-$) if all the edges are taken outgoing from (incoming to) the partition defined by $\lambda_p$. 

%++++++++++++++++++++++++++++++++++++++++++++++++++++++++++++++++++++++++++++++++++++
\begin{figure}[h!]
    \centering
    \includegraphics[width=0.80\linewidth]{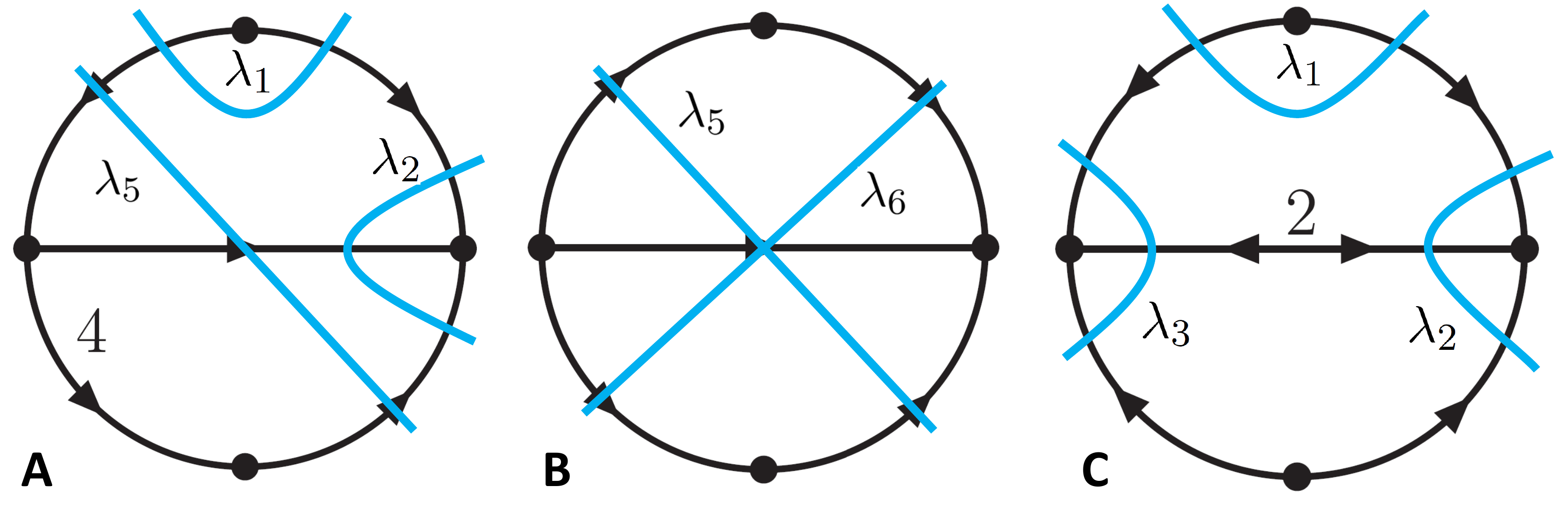}
    \caption{Graphical interpretation from causal geometrical selection rules of the two-eloop topology from Fig. \ref{fig:TwoLoopsQC}. Although all configurations depicted are acyclic, the selected thresholds are not causally entangled because: (A) not all the edges are cut and the momentum flows in $\lambda_5$ are  not aligned in the same direction; (B) $\lambda_5$ and $\lambda_6$ cross each other; and (C) the momentum of $e_2$ can not be consistently oriented.}
    \label{fig:AppA}
\end{figure}
%++++++++++++++++++++++++++++++++++++++++++++++++++++++++++++++++++++++++++++++++++++

The second step corresponds to the determination of the causal entangled thresholds. Since this diagram has $V=4$, it is of order $k=V-1=3$, that implies that the causal representation is obtained by multiplying 3 causal propagators. Equivalently, it means that the diagram can be decomposed by \emph{simultaneously imposing} 3 different causal cuts. In principle, there are 20 possibilities but the conditions 1-3 listed in Sec. \ref{sec:Causalflow} limit them to 10. Explicitly, we have
\begin{align}
    \Sigma = 
    \big\{ &\{1,2,4\},\{1,2,6\},\{1,3,4\},\{1,3,5\},\{1,4,5\}, \nn\\
    &\{1,4,6\},\{2,3,5\},\{2,3,6\},\{2,4,5\},\{3,4,6\}  \big\} ~,
\end{align}
where we use the simplified notation $\lambda_i \lambda_j \lambda_k \equiv \{i,j,k\}$. Examples of partitions that are not causally entangled are:
\begin{enumerate}
    \item $\{1,2,5\}$ is not allowed because $e_4$ remains uncut, i.e. condition 1 is not fulfilled.
    \item entangled thresholds containing $\{5,6\}$ are not allowed because $\lambda_5$ and $\lambda_6$ cross each other, i.e. condition 2 is not fulfilled.
    \item $\{1,2,3\}$ is not allowed because $e_2$ can not be oriented in a consistent way. Explicitly, if we take $\lambda_1^+$, then we have $\lambda_2^-$ and $\lambda_3^-$; but $\lambda_2^-$ implies $S(e_2,G_0)=0$ and $\lambda_3^-$ implies $S(e_2,G_0)=1$ (following the notation introduced in Sec. \ref{sec:Hamiltonian}).
\end{enumerate}
These 3 cases are depicted in Fig. \ref{fig:AppA}. Finally, we would like to close this Appendix by reminding that conditions 1-3 are equivalent to the following ones:
\begin{enumerate}
 \item The edges must be oriented in such a way that cyclic loops are not present, i.e. we need to identify all the possible directed acyclic graphs obtained from the original reduced Feynman graph.
 \item Then, we need to dress the acyclic graphs with $k=V-1$ simultaneous causal propagators, in such a way that the conditions 1-2 from the list in Sec. \ref{sec:Causalflow} are fulfilled.
\end{enumerate}
This was the strategy first used in Ref. \cite{Ramirez-Uribe:2021ubp} to bootstrap the causal representation, and justifies the importance of developing an efficient algorithm to detect directed acyclic configurations for a given topology.

%%%%%%%%%%%%%%%%%%%%%%%%%%%%%%%%%%%%%%%%%%%%%%%%%%%%%%%%%%%%%%%%%%%%%%%%%%%%%%
%%%%%%%%%%%%%%%%%%%%%%%%%%%%%%%%%%%%%%%%%%%%%%%%%%%%%%%%%%%%%%%%%%%%%%%%%%%%%%
\section{Compact encoding of vertex registers}
\label{app:Encoding}
Having in mind the extension of the VQE algorithm presented in this paper to complex multiloop topologies, it is relevant to think about suitable scaling techniques. As we already know from Ref.~\cite{Ramirez-Uribe:2021ubp}, the number of qubits required to implement the computation scales very fast with the complexity of the underlying graphs. For this, it is important to identify strategies that allow us to reduce the qubit consumption. 

One source of proliferation of qubits is associated with the description of the vertex space $V$. The encoding used for the vertex registers in the discussion above is of the one-hot kind, i.e., each classical vertex state for a specific vertex with label $v\in \{0,\dots,|V|-1\}$ is associated to a unique string with all $0$'s except for a single $1$ on the $v$-th position. For example, the vertices in Fig.~\ref{fig:TwoLoopsQC} are associated to the following elements of the computational basis: 
\begin{equation}
\begin{aligned}
    v_0 &\mapsto \ket{0001}_V;\\ 
    v_1 &\mapsto \ket{0010}_V;\\
    v_2 &\mapsto \ket{0100}_V;\\
    v_3 &\mapsto \ket{1000}_V.
\end{aligned}
\end{equation}
With this encoding, the oriented link $e_{(i,j)} : v_i \rightarrow v_j$ is represented by an hopping term $\sigma^-_{v_i}\sigma^+_{v_j} \pi^0_{e_{(i,j)}}$.
However, it is simpler (and more scalable) to consider a compact encoding using $q=\lceil \log_2(|V|)\rceil$ qubits instead of the $|V|$ necessary in the one-hot case\footnote{Depending on the problem at hand, other compact encoding like the Gray~\cite{GRAYCODE} one could be useful. However, a careful study of the total efficiency should be carried out, since this encoding might be harder to decode. For this reason, here we restrict our attention to the binary and one-hot encoding only.}. Returning to the example in Fig.~\ref{fig:TwoLoopsQC}, one can use a possible compact encoding with just $q=2$ qubits, as the following:
\begin{equation}
\begin{aligned}
    v_0 &\mapsto \ket{00}_V;\\ 
    v_1 &\mapsto \ket{01}_V;\\
    v_2 &\mapsto \ket{10}_V;\\
    v_3 &\mapsto \ket{11}_V.
\end{aligned}
\end{equation}
The hopping terms for the adjacency matrix can be computed, as for the one-hot encoding, by looking at the oriented link $e_{(i,j)} : v_i \rightarrow v_j$, which will contain a mix of ladder (at least one) and projector operators. Using Fig.~\ref{fig:TwoLoopsQC}, the oriented link $e_{(1,0)}: v_1 \rightarrow v_0$ 
is associated with the non-Hermitian operator 
\beq 
\ketbra{00}{01}_V \otimes \pi^0_{e_{(1,0)}} = 
{(\sigma^-_{1} \otimes \sigma^+_{0})}_V \otimes \pi^0_{e_{(1,0)}} \, .
\eeq 
After computing all the terms for the operator representing the adjacency matrix $A$, one can build the loop Hamiltonian by tracing out the vertex register as in Eqs.~\eqref{eq:loopHam_trn}-\eqref{eq:loopHam_treA} (or any equivalent positive semi-definite matrix with $\widetilde{\mathcal{H}}_E$ as kernel). This is guaranteed by the fact that the only terms contributing to $H_G$ after computing the trace are those in $A^n$ that do not contain ladder operators (i.e., diagonal terms), because these can be obtained only if the products of the $n$ hopping terms connect the same vertex as input (e.g., $\ketbra{v_i}{v_i}_V$). In other words, the terms present in $A^n$ are those representing a closed chain, starting and ending in the same vertex.

%%%%%%%%%%%%%%%%%%%%%%%%%%%%%%%%%%%%%%%%%%%%%%%%%%%%%%%%%%%%%%%%%%%%%%%%%%%%%%
%%%%%%%%%%%%%%%%%%%%%%%%%%%%%%%%%%%%%%%%%%%%%%%%%%%%%%%%%%%%%%%%%%%%%%%%%%%%%%
\section{Loop Hamiltonians for representative multiloop topologies}
\label{app:Hamiltonians}
In this appendix, we include the formulae for the loop Hamiltonians for the multiloop topologies described in Fig. \ref{fig:digraph6}. Results for diagram A were discussed in detail in Sec. \ref{ssec:2eloopHam}, so here we present only expressions for the remaining topologies. Explicitly, we have for diagram B (three eloops with four vertices, or Mercedes diagram), 
%\beqn
%\nn H_{\rm Top. B} &=& 3 \pi_{0}^0 \,  \pi_{1}^0 \,  \pi_{2}^0 \,  + 3 \pi_{0}^1 \,  \pi_{1}^1 \,  \pi_{2}^1 \,  + 3 \pi_{0}^1 \,  \pi_{3}^1 \,  \pi_{4}^0 \,  + 4 \pi_{1}^0 \,  \pi_{2}^0 \,  \pi_{3}^1 \,  \pi_{4}^0 \,  + 3 \pi_{0}^0 \,  \pi_{3}^0 \,  \pi_{4}^1 \,  + 4 \pi_{1}^1 \,  \pi_{2}^1 \,  \pi_{3}^0 \,  \pi_{4}^1 \,  
%\\ \nn &+& 4 \pi_{0}^1 \,  \pi_{1}^1 \,  \pi_{3}^1 \,  \pi_{5}^0 \,  + 3 \pi_{2}^0 \,  \pi_{3}^1 \,  \pi_{5}^0 \,  + 3 \pi_{1}^1 \,  \pi_{4}^1 \,  \pi_{5}^0 \,  + 4 \pi_{0}^0 \,  \pi_{2}^0 \,  \pi_{4}^1 \,  \pi_{5}^0 \,  + 4 \pi_{0}^0 \,  \pi_{1}^0 \,  \pi_{3}^0 \,  \pi_{5}^1 \,  
%\\ &+& 3 \pi_{2}^1 \,  \pi_{3}^0 \,  \pi_{5}^1 \,  + 3 \pi_{1}^0 \,  \pi_{4}^0 \,  \pi_{5}^1 \,  + 4 \pi_{0}^1 \,  \pi_{2}^1 \,  \pi_{4}^0 \,  \pi_{5}^1 \,   ,
%\label{eq:HamiltonianoB}
%\eeqn
\begin{align}\label{eq:HamiltonianoB}
    H_{\rm Top. B} 
    &= 3 \pi_{0}^0 \,  \pi_{1}^0 \,  \pi_{2}^0 \,  + 3 \pi_{0}^1 \,  \pi_{1}^1 \,  \pi_{2}^1 \,  
    + 3 \pi_{0}^1 \,  \pi_{3}^1 \,  \pi_{4}^0 \, \nn\\
    &+ 4 \pi_{1}^0 \,  \pi_{2}^0 \,  \pi_{3}^1 \,  \pi_{4}^0 \,  
    + 3 \pi_{0}^0 \,  \pi_{3}^0 \,  \pi_{4}^1 \,   + 4 \pi_{1}^1 \,  \pi_{2}^1 \,  \pi_{3}^0 \,  \pi_{4}^1 \, \nn\\ 
    &+ 4 \pi_{0}^1 \,  \pi_{1}^1 \,  \pi_{3}^1 \,  \pi_{5}^0 \,  + 3 \pi_{2}^0 \,  \pi_{3}^1 \,  \pi_{5}^0 \,  
    + 3 \pi_{1}^1 \,  \pi_{4}^1 \,  \pi_{5}^0 \,  \nn\\
    &+ 4 \pi_{0}^0 \,  \pi_{2}^0 \,  \pi_{4}^1 \,  \pi_{5}^0 \,  
    + 4 \pi_{0}^0 \,  \pi_{1}^0 \,  \pi_{3}^0 \,  \pi_{5}^1 \,  + 3 \pi_{2}^1 \,  \pi_{3}^0 \,  \pi_{5}^1 \,  \nn\\
    &+ 3 \pi_{1}^0 \,  \pi_{4}^0 \,  \pi_{5}^1 \,  + 4 \pi_{0}^1 \,  \pi_{2}^1 \,  \pi_{4}^0 \,  \pi_{5}^1 \,   ,
\end{align}
for diagram C (four eloops with five vertices, or pizza diagram),
%\beqn
%\nn H_{\rm Top. C} &=& 4\pi_{0}^0\,  \pi_{1}^0\,  \pi_{2}^0\,  \pi_{3}^0\,+4\pi_{0}^1\,  \pi_{1}^1\,  \pi_{2}^1\,  \pi_{3}^1\,+3\pi_{0}^1\,  \pi_{4}^1\,  \pi_{5}^0\,+5\pi_{1}^0\,  \pi_{2}^0\,  \pi_{3}^0\,  \pi_{4}^1\,  \pi_{5}^0\,+3\pi_{0}^0\,  \pi_{4}^0\,  \pi_{5}^1\,
%\\ \nn &+& 5\pi_{1}^1\,  \pi_{2}^1\,  \pi_{3}^1\,  \pi_{4}^0\,  \pi_{5}^1\,+4\pi_{0}^1\,  \pi_{1}^1\,  \pi_{4}^1\,  \pi_{6}^0\,+4\pi_{2}^0\,  \pi_{3}^0\,  \pi_{4}^1\,  \pi_{6}^0\,+3\pi_{1}^1\,  \pi_{5}^1\,  \pi_{6}^0\,+5\pi_{0}^0\,  \pi_{2}^0\,  \pi_{3}^0\,  \pi_{5}^1\,  \pi_{6}^0\,
%\\ \nn &+& 4\pi_{0}^0\,  \pi_{1}^0\,  \pi_{4}^0\,  \pi_{6}^1\,+4\pi_{2}^1\,  \pi_{3}^1\,  \pi_{4}^0\,  \pi_{6}^1\,+3\pi_{1}^0\,  \pi_{5}^0\,  \pi_{6}^1\,+5\pi_{0}^1\,  \pi_{2}^1\,  \pi_{3}^1\,  \pi_{5}^0\,  \pi_{6}^1\,+5\pi_{0}^1\,  \pi_{1}^1\,  \pi_{2}^1\,  \pi_{4}^1\,  \pi_{7}^0\,
%\\ \nn&+& 3\pi_{3}^0\,  \pi_{4}^1\,  \pi_{7}^0\,+4\pi_{1}^1\,  \pi_{2}^1\,  \pi_{5}^1\,  \pi_{7}^0\,+4\pi_{0}^0\,  \pi_{3}^0\,  \pi_{5}^1\,  \pi_{7}^0\,+3\pi_{2}^1\,  \pi_{6}^1\,  \pi_{7}^0\,+5\pi_{0}^0\,  \pi_{1}^0\,  \pi_{3}^0\,  \pi_{6}^1\,  \pi_{7}^0\,
%\\ \nn&+& 5\pi_{0}^0\,  \pi_{1}^0\,  \pi_{2}^0\,  \pi_{4}^0\,  \pi_{7}^1\,+3\pi_{3}^1\,  \pi_{4}^0\,  \pi_{7}^1\,+4\pi_{1}^0\,  \pi_{2}^0\,  \pi_{5}^0\,  \pi_{7}^1\,+4\pi_{0}^1\,  \pi_{3}^1\,  \pi_{5}^0\,  \pi_{7}^1\,+3\pi_{2}^0\,  \pi_{6}^0\,  \pi_{7}^1\,
%\\ &+& 5\pi_{0}^1\,  \pi_{1}^1\,  \pi_{3}^1\,  \pi_{6}^0\,  \pi_{7}^1 \, ,
%\label{eq:HamiltonianoC}
%\eeqn
\begin{align}
    H_{\rm Top. C} 
    &=4\pi_{0}^0\,  \pi_{1}^0\,  \pi_{2}^0\,  \pi_{3}^0\,
    + 4\pi_{0}^1\,  \pi_{1}^1\,  \pi_{2}^1\,  \pi_{3}^1\,
    + 3\pi_{0}^1\,  \pi_{4}^1\,  \pi_{5}^0\, \nn\\
    &+5\pi_{1}^0\,  \pi_{2}^0\,  \pi_{3}^0\,  \pi_{4}^1\,  \pi_{5}^0\,
    + 3\pi_{0}^0\,  \pi_{4}^0\,  \pi_{5}^1\,
    + 5\pi_{1}^1\,  \pi_{2}^1\,  \pi_{3}^1\,  \pi_{4}^0\,  \pi_{5}^1\, \nn\\
    &+4\pi_{0}^1\,  \pi_{1}^1\,  \pi_{4}^1\,  \pi_{6}^0\,
    + 4\pi_{2}^0\,  \pi_{3}^0\,  \pi_{4}^1\,  \pi_{6}^0\,
    + 3\pi_{1}^1\,  \pi_{5}^1\,  \pi_{6}^0\, \nn\\
    &+5\pi_{0}^0\,  \pi_{2}^0\,  \pi_{3}^0\,  \pi_{5}^1\,  \pi_{6}^0\,
    + 4\pi_{0}^0\,  \pi_{1}^0\,  \pi_{4}^0\,  \pi_{6}^1\,
    + 4\pi_{2}^1\,  \pi_{3}^1\,  \pi_{4}^0\,  \pi_{6}^1\, \nn\\
    &+ 3\pi_{1}^0\,  \pi_{5}^0\,  \pi_{6}^1\,
    + 5\pi_{0}^1\,  \pi_{2}^1\,  \pi_{3}^1\,  \pi_{5}^0\,  \pi_{6}^1\,
    + 5\pi_{0}^1\,  \pi_{1}^1\,  \pi_{2}^1\,  \pi_{4}^1\,  \pi_{7}^0\, \nn\\
    &+ 3\pi_{3}^0\,  \pi_{4}^1\,  \pi_{7}^0\,
    + 4\pi_{1}^1\,  \pi_{2}^1\,  \pi_{5}^1\,  \pi_{7}^0\,
    + 4\pi_{0}^0\,  \pi_{3}^0\,  \pi_{5}^1\,  \pi_{7}^0\, \nn\\
    &+ 3\pi_{2}^1\,  \pi_{6}^1\,  \pi_{7}^0\,
    + 5\pi_{0}^0\,  \pi_{1}^0\,  \pi_{3}^0\,  \pi_{6}^1\,  \pi_{7}^0\,
    + 5\pi_{0}^0\,  \pi_{1}^0\,  \pi_{2}^0\,  \pi_{4}^0\,  \pi_{7}^1\, \nn\\
    &+ 3\pi_{3}^1\,  \pi_{4}^0\,  \pi_{7}^1\,
    + 4\pi_{1}^0\,  \pi_{2}^0\,  \pi_{5}^0\,  \pi_{7}^1\,
    + 4\pi_{0}^1\,  \pi_{3}^1\,  \pi_{5}^0\,  \pi_{7}^1\, \nn\\
    &+ 3\pi_{2}^0\,  \pi_{6}^0\,  \pi_{7}^1\,
    + 5\pi_{0}^1\,  \pi_{1}^1\,  \pi_{3}^1\,  \pi_{6}^0\,  \pi_{7}^1 \, ,
\label{eq:HamiltonianoC}
\end{align}
for diagram D (four eloops with six-vertices in $t$-channel configuration),
\begin{align}
    &H_{\rm Top. D} 
     =  4 \pi_{0}^0 \,  \pi_{1}^0 \,  \pi_{2}^0 \,  \pi_{3}^0 \,  
     +  4 \pi_{0}^1 \,  \pi_{1}^1 \,  \pi_{2}^1 \,  \pi_{3}^1 \,  
     +  6 \pi_{1}^1 \,  \pi_{5}^1 \,  \pi_{6}^0 \,  
\nn\\&+ 5 \pi_{0}^0 \,  \pi_{2}^0 \,  \pi_{3}^0 \,  \pi_{5}^1 \,  \pi_{6}^0 \,  
     +  6 \pi_{1}^0 \,  \pi_{5}^0 \,  \pi_{6}^1 \,  
     +  5 \pi_{0}^1 \,  \pi_{2}^1 \,  \pi_{3}^1 \,  \pi_{5}^0 \,  \pi_{6}^1 \,  
\nn\\&+ 5 \pi_{0}^1 \,  \pi_{1}^1 \,  \pi_{2}^1 \,  \pi_{4}^1 \,  \pi_{7}^0 \,  
     +  6 \pi_{3}^0 \,  \pi_{4}^1 \,  \pi_{7}^0 \,  
     +  6 \pi_{0}^1 \,  \pi_{2}^1 \,  \pi_{4}^1 \,  \pi_{5}^0 \,  \pi_{6}^1 \,  \pi_{7}^0 \,  
\nn\\&+ 5 \pi_{0}^0 \,  \pi_{1}^0 \,  \pi_{2}^0 \,  \pi_{4}^0 \,  \pi_{7}^1 \,  
     +  6 \pi_{3}^1 \,  \pi_{4}^0 \,  \pi_{7}^1 \,  
     +  6 \pi_{0}^0 \,  \pi_{2}^0 \,  \pi_{4}^0 \,  \pi_{5}^1 \,  \pi_{6}^0 \,  \pi_{7}^1 \,  
\nn\\&+ 4 \pi_{0}^1 \,  \pi_{4}^1 \,  \pi_{5}^0 \,  \pi_{8}^0 \,  
     +  6 \pi_{1}^0 \,  \pi_{2}^0 \,  \pi_{3}^0 \,  \pi_{4}^1 \,  \pi_{5}^0 \,  \pi_{8}^0 \,  
     +  5 \pi_{0}^1 \,  \pi_{1}^1 \,  \pi_{4}^1 \,  \pi_{6}^0 \,  \pi_{8}^0 \,  
\nn\\&+ 5 \pi_{2}^0 \,  \pi_{3}^0 \,  \pi_{4}^1 \,  \pi_{6}^0 \,  \pi_{8}^0 \,  
     +  5 \pi_{1}^0 \,  \pi_{2}^0 \,  \pi_{5}^0 \,  \pi_{7}^1 \,  \pi_{8}^0 \,  
     +  5 \pi_{0}^1 \,  \pi_{3}^1 \,  \pi_{5}^0 \,  \pi_{7}^1 \,  \pi_{8}^0 \,  
\nn\\&+ 4 \pi_{2}^0 \,  \pi_{6}^0 \,  \pi_{7}^1 \,  \pi_{8}^0 \,  
     +  6 \pi_{0}^1 \,  \pi_{1}^1 \,  \pi_{3}^1 \,  \pi_{6}^0 \,  \pi_{7}^1 \,  \pi_{8}^0 \,  
     +  4 \pi_{0}^0 \,  \pi_{4}^0 \,  \pi_{5}^1 \,  \pi_{8}^1 \,  
\nn\\&+ 6 \pi_{1}^1 \,  \pi_{2}^1 \,  \pi_{3}^1 \,  \pi_{4}^0 \,  \pi_{5}^1 \,  \pi_{8}^1 \,  
     +  5 \pi_{0}^0 \,  \pi_{1}^0 \,  \pi_{4}^0 \,  \pi_{6}^1 \,  \pi_{8}^1 \,  
     +  5 \pi_{2}^1 \,  \pi_{3}^1 \,  \pi_{4}^0 \,  \pi_{6}^1 \,  \pi_{8}^1 \,  
\nn\\&+ 5 \pi_{1}^1 \,  \pi_{2}^1 \,  \pi_{5}^1 \,  \pi_{7}^0 \,  \pi_{8}^1 \,  
     +  5 \pi_{0}^0 \,  \pi_{3}^0 \,  \pi_{5}^1 \,  \pi_{7}^0 \,  \pi_{8}^1 \,  
     +  4 \pi_{2}^1 \,  \pi_{6}^1 \,  \pi_{7}^0 \,  \pi_{8}^1 \,  
\nn\\&+ 6 \pi_{0}^0 \,  \pi_{1}^0 \,  \pi_{3}^0 \,  \pi_{6}^1 \,  \pi_{7}^0 \,  \pi_{8}^1 \,   ,
\label{eq:HamiltonianoD}
\end{align}
for diagram E (four eloops with six-vertices in $s$-channel configuration),
\begin{align}
&H_{\rm Top. E} 
      = 4 \pi_{0}^0 \,  \pi_{1}^0 \,  \pi_{2}^0 \,  \pi_{3}^0 \,  
      + 4 \pi_{0}^1 \,  \pi_{1}^1 \,  \pi_{2}^1 \,  \pi_{3}^1 \,  
      + 6 \pi_{0}^1 \,  \pi_{4}^1 \,  \pi_{5}^0 \,  
\nn\\&+ 5 \pi_{1}^0 \,  \pi_{2}^0 \,  \pi_{3}^0 \,  \pi_{4}^1 \,  \pi_{5}^0 \,  
      + 6 \pi_{0}^0 \,  \pi_{4}^0 \,  \pi_{5}^1 \,  
      + 5 \pi_{1}^1 \,  \pi_{2}^1 \,  \pi_{3}^1 \,  \pi_{4}^0 \,  \pi_{5}^1 \,  
\nn\\&+ 6 \pi_{2}^1 \,  \pi_{6}^1 \,  \pi_{7}^0 \,  
      + 5 \pi_{0}^0 \,  \pi_{1}^0 \,  \pi_{3}^0 \,  \pi_{6}^1 \,  \pi_{7}^0 \,  
      + 6 \pi_{1}^0 \,  \pi_{3}^0 \,  \pi_{4}^1 \,  \pi_{5}^0 \,  \pi_{6}^1 \,  \pi_{7}^0 \,  
\nn\\&+ 6 \pi_{2}^0 \,  \pi_{6}^0 \,  \pi_{7}^1 \,  
      + 5 \pi_{0}^1 \,  \pi_{1}^1 \,  \pi_{3}^1 \,  \pi_{6}^0 \,  \pi_{7}^1 \,  
      + 6 \pi_{1}^1 \,  \pi_{3}^1 \,  \pi_{4}^0 \,  \pi_{5}^1 \,  \pi_{6}^0 \,  \pi_{7}^1 \,  
\nn\\&+ 5 \pi_{0}^0 \,  \pi_{1}^0 \,  \pi_{4}^0 \,  \pi_{6}^1 \,  \pi_{8}^0 \,  
      + 5 \pi_{2}^1 \,  \pi_{3}^1 \,  \pi_{4}^0 \,  \pi_{6}^1 \,  \pi_{8}^0 \,  
      + 4 \pi_{1}^0 \,  \pi_{5}^0 \,  \pi_{6}^1 \,  \pi_{8}^0 \,  
\nn\\&+ 6 \pi_{0}^1 \,  \pi_{2}^1 \,  \pi_{3}^1 \,  \pi_{5}^0 \,  \pi_{6}^1 \,  \pi_{8}^0 \,  
      + 6 \pi_{0}^0 \,  \pi_{1}^0 \,  \pi_{2}^0 \,  \pi_{4}^0 \,  \pi_{7}^1 \,  \pi_{8}^0 \,  
      + 4 \pi_{3}^1 \,  \pi_{4}^0 \,  \pi_{7}^1 \,  \pi_{8}^0 \,  
\nn\\&+ 5 \pi_{1}^0 \,  \pi_{2}^0 \,  \pi_{5}^0 \,  \pi_{7}^1 \,  \pi_{8}^0 \,  
      + 5 \pi_{0}^1 \,  \pi_{3}^1 \,  \pi_{5}^0 \,  \pi_{7}^1 \,  \pi_{8}^0 \,  
      + 5 \pi_{0}^1 \,  \pi_{1}^1 \,  \pi_{4}^1 \,  \pi_{6}^0 \,  \pi_{8}^1 \,  
\nn\\&+ 5 \pi_{2}^0 \,  \pi_{3}^0 \,  \pi_{4}^1 \,  \pi_{6}^0 \,  \pi_{8}^1 \,  
      + 4 \pi_{1}^1 \,  \pi_{5}^1 \,  \pi_{6}^0 \,  \pi_{8}^1 \,  
      + 6 \pi_{0}^0 \,  \pi_{2}^0 \,  \pi_{3}^0 \,  \pi_{5}^1 \,  \pi_{6}^0 \,  \pi_{8}^1 \,  
\nn\\&+ 6 \pi_{0}^1 \,  \pi_{1}^1 \,  \pi_{2}^1 \,  \pi_{4}^1 \,  \pi_{7}^0 \,  \pi_{8}^1 \,  
      + 4 \pi_{3}^0 \,  \pi_{4}^1 \,  \pi_{7}^0 \,  \pi_{8}^1 \,  
      + 5 \pi_{1}^1 \,  \pi_{2}^1 \,  \pi_{5}^1 \,  \pi_{7}^0 \,  \pi_{8}^1 \,  
\nn\\&+ 5 \pi_{0}^0 \,  \pi_{3}^0 \,  \pi_{5}^1 \,  \pi_{7}^0 \,  \pi_{8}^1 \,   ,
\label{eq:HamiltonianoE}
\end{align}
and for diagram F (four eloops with six-vertices in $u$-channel configuration),
\begin{align}
&H_{\rm Top. F} 
        = 4 \pi_{0}^0 \,  \pi_{1}^0 \,  \pi_{2}^0 \,  \pi_{3}^0 \,  
        + 4 \pi_{0}^1 \,  \pi_{1}^1 \,  \pi_{2}^1 \,  \pi_{3}^1 \,  
        + 4 \pi_{0}^1 \,  \pi_{1}^1 \,  \pi_{4}^1 \,  \pi_{6}^0 \,  
\nn\\&  + 4 \pi_{2}^0 \,  \pi_{3}^0 \,  \pi_{4}^1 \,  \pi_{6}^0 \,  
        + 4 \pi_{0}^0 \,  \pi_{1}^0 \,  \pi_{4}^0 \,  \pi_{6}^1 \,  
        + 4 \pi_{2}^1 \,  \pi_{3}^1 \,  \pi_{4}^0 \,  \pi_{6}^1 \,  
\nn\\&  + 4 \pi_{1}^1 \,  \pi_{2}^1 \,  \pi_{5}^1 \,  \pi_{7}^0 \,  
        + 4 \pi_{0}^0 \,  \pi_{3}^0 \,  \pi_{5}^1 \,  \pi_{7}^0 \,  
        + 6 \pi_{1}^1 \,  \pi_{3}^0 \,  \pi_{4}^1 \,  \pi_{5}^1 \,  \pi_{6}^0 \,  \pi_{7}^0 \,  
\nn\\&  + 6 \pi_{0}^0 \,  \pi_{2}^1 \,  \pi_{4}^0 \,  \pi_{5}^1 \,  \pi_{6}^1 \,  \pi_{7}^0 \,  
        + 4 \pi_{1}^0 \,  \pi_{2}^0 \,  \pi_{5}^0 \,  \pi_{7}^1 \,  
        + 4 \pi_{0}^1 \,  \pi_{3}^1 \,  \pi_{5}^0 \,  \pi_{7}^1 \,  
\nn\\&  + 6 \pi_{0}^1 \,  \pi_{2}^0 \,  \pi_{4}^1 \,  \pi_{5}^0 \,  \pi_{6}^0 \,  \pi_{7}^1 \,  
        + 6 \pi_{1}^0 \,  \pi_{3}^1 \,  \pi_{4}^0 \,  \pi_{5}^0 \,  \pi_{6}^1 \,  \pi_{7}^1 \,  
        + 4 \pi_{0}^1 \,  \pi_{4}^1 \,  \pi_{5}^0 \,  \pi_{8}^0 \,  
\nn\\&  + 6 \pi_{1}^0 \,  \pi_{2}^0 \,  \pi_{3}^0 \,  \pi_{4}^1 \,  \pi_{5}^0 \,  \pi_{8}^0 \,  
        + 4 \pi_{1}^0 \,  \pi_{5}^0 \,  \pi_{6}^1 \,  \pi_{8}^0 \,  
        + 6 \pi_{0}^1 \,  \pi_{2}^1 \,  \pi_{3}^1 \,  \pi_{5}^0 \,  \pi_{6}^1 \,  \pi_{8}^0 \,  
\nn\\&  + 6 \pi_{0}^1 \,  \pi_{1}^1 \,  \pi_{2}^1 \,  \pi_{4}^1 \,  \pi_{7}^0 \,  \pi_{8}^0 \,  
        + 4 \pi_{3}^0 \,  \pi_{4}^1 \,  \pi_{7}^0 \,  \pi_{8}^0 \,  
        + 4 \pi_{2}^1 \,  \pi_{6}^1 \,  \pi_{7}^0 \,  \pi_{8}^0 \,  
\nn\\&  + 6 \pi_{0}^0 \,  \pi_{1}^0 \,  \pi_{3}^0 \,  \pi_{6}^1 \,  \pi_{7}^0 \,  \pi_{8}^0 \,  
        + 4 \pi_{0}^0 \,  \pi_{4}^0 \,  \pi_{5}^1 \,  \pi_{8}^1 \,  
        + 6 \pi_{1}^1 \,  \pi_{2}^1 \,  \pi_{3}^1 \,  \pi_{4}^0 \,  \pi_{5}^1 \,  \pi_{8}^1 \,  
\nn\\&  + 4 \pi_{1}^1 \,  \pi_{5}^1 \,  \pi_{6}^0 \,  \pi_{8}^1 \,  
        + 6 \pi_{0}^0 \,  \pi_{2}^0 \,  \pi_{3}^0 \,  \pi_{5}^1 \,  \pi_{6}^0 \,  \pi_{8}^1 \,  
        + 6 \pi_{0}^0 \,  \pi_{1}^0 \,  \pi_{2}^0 \,  \pi_{4}^0 \,  \pi_{7}^1 \,  \pi_{8}^1 \,  
\nn\\&  + 4 \pi_{3}^1 \,  \pi_{4}^0 \,  \pi_{7}^1 \,  \pi_{8}^1 \,  
        + 4 \pi_{2}^0 \,  \pi_{6}^0 \,  \pi_{7}^1 \,  \pi_{8}^1 \, 
        + 6 \pi_{0}^1 \,  \pi_{1}^1 \,  \pi_{3}^1 \,  \pi_{6}^0 \,  \pi_{7}^1 \,  \pi_{8}^1 \, .
\label{eq:HamiltonianoF}
\end{align}
We followed the approach described in Sec.~\ref{sec:Hamiltonian}, in particular using \Eq{eq:loopHam_trn}.

%%%%%%%%%%%%%%%%%%%%%%%%%%%%%%%%%%%%%%%%%%%%%%%%%%%%%%%%%%%%%%%%%%%%%%%%%%%%%%%%%
%\bibliography{refs}

%%%%%%%%%%%%%%%%%%%%%%%%%%%%%%%%%%%%%%%%%%%%%%%%%%%%%%%%%%%%%%%%%
%%%%%%%%%%%%%%%%%%%%%%%%%%%%%%%%%%%%%%%%%%%%%%%%%%%%%%%%%%%%%%%%%
\end{document}